\documentclass[10pt,preprint2]{aastex}
\newcommand{\V}[1]{\mathbf{#1}} 
\newcommand\Alfven{Alfv\'en } 
\newcommand\Alfvenic{Alfv\'enic } 
\usepackage{fullpage}
\usepackage{amssymb}
\usepackage{amsmath}
\usepackage{hyperref}
\usepackage{natbib}
\usepackage{fullpage}
\usepackage{color}

\begin{document}
\title{
Physical Interpretation of the Angle Dependent Magnetic Helicity
Spectrum in the Solar Wind: The Nature of Turbulent Fluctuations near
the Proton Gyroradius Scale
}
\author{Kristopher~G. Klein, Gregory~G. Howes}
\affil{Department of Physics and Astronomy, 
University of Iowa, Iowa City, 
Iowa 52242, USA.}
\email{kristopher-klein@uiowa.edu}

\and

\author{
Jason~M. TenBarge
}
\affil{
IREAP, University of Maryland, College Park, Maryland 20742, USA.
}

\and

\author{John~J. Podesta}
\affil{Center for Space Plasma Physics, Space Science Institute,
Boulder, CO 80301}

\begin{abstract}
Motivated by recent observations of distinct parallel and perpendicular signatures
in magnetic helicity measurements segregated by wave period and angle between 
the local magnetic field and the solar wind velocity, this paper undertakes a 
comparison of three intervals of 
\emph{Ulysses} data with synthetic time series generated from a physically motivated 
turbulence model. From these comparisons, it is hypothesized that the
observed signatures result from a perpendicular cascade of \Alfvenic fluctuations
and a local, non-turbulent population of ion cyclotron or whistler waves generated
by temperature anisotropy instabilities. By constraining the model's free parameters
through comparison to \emph{in situ} data,
it is found that, on average, $ \sim 95\%$ of the power near 
dissipative scales is contained in a
perpendicular \Alfvenic cascade and that the parallel fluctuations 
are propagating nearly unidirectionally. The effects of aliasing on magnetic helicity
measurements are considered and shown to be significant near the Nyquist frequency.
\end{abstract}




\section{Introduction}
\label{sec:Intro}




Turbulence is ubiquitous in space and astrophysical plasmas.
In the heliosphere, it plays a critical role in plasma heating
and the scattering of cosmic rays and energetic solar particles.
Given the availability of
detailed \emph{in situ} measurements from a broad range of
heliophysics missions, the near-Earth solar wind is an ideal
environment in which to study the fundamental nature of plasma
turbulence.  In the solar wind, the primary focus is the transport of
energy from large scale turbulent motions to the small length scales
at which the turbulence may be dissipated and the energy ultimately
converted to plasma heat.  Key questions at the forefront of solar
wind turbulence investigations are: (i) what are the characteristics
of the turbulent fluctuations?; (ii) what are the properties of the
nonlinear interactions that drive the turbulent cascade of energy?;
and (iii) what are the physical dissipation mechanisms that damp the
turbulence and ultimately lead to plasma heating? The answers to each
of these questions are interdependent.  In particular, the nature of
the electromagnetic and plasma fluctuations underlying the turbulence
constrains the possible dissipation mechanisms.

The definitive determination of the characteristics of the turbulent
fluctuations, however, is made difficult due to the fact that the bulk
of current \emph{in situ} solar wind observations are single point
measurements. The Taylor hypothesis \citep{Taylor:1938} is typically
invoked, because of the super-\Alfvenic velocity of the solar wind
near 1 AU, to transform from the observed frequency of the advected turbulent
fluctuations to a length scale of the fluctuations. 
For low frequency fluctuations with $\omega \lesssim \Omega_i$ 
and with the exception
of a small number of studies of solar wind turbulence employing
analyses based on multi-spacecraft measurements, we have no knowledge
of the frequency of the fluctuations in the rest frame of the solar wind
plasma. This limitation drives the quest to exploit alternative means
to illuminate the character of the turbulent fluctuations in the solar
wind. A wide range of characteristics, including polarizations,
helicities, and other transport ratios have previously been used to
decipher the dynamics of space plasmas \citep{Song:1994,Gary:1992,
Lacombe:1995}. Here we employ the magnetic helicity as a sensitive
probe for turbulent fluctuations in the solar wind.

The \emph{fluctuating magnetic helicity}, defined as $H_m'\equiv \int
d^3\V{r}\ 
\delta \V{A}
\cdot \delta\V{B}$, where $\delta \V{A}$ and $\delta \V{B}$ are
respectively the fluctuations of the vector potential and magnetic
field, was first proposed as a useful metric for studying turbulent
fluctuations in the solar wind by \cite{Matthaeus:1982b}. The \emph{reduced
fluctuating magnetic helicity}, $H_m ^{r'}$ \citep{Matthaeus:1982a}, is
related to $H_m'$ but is derivable from single-point spacecraft
measurements and is a function of  spacecraft-frame frequency
$\omega_{SC}$,
\citep{Howes:2010a}:
\begin{equation}
\begin{split}
H_m^{r'}(\omega_{SC})=&\sum_{\V{k}} \frac{i[\V{B}_2(\V{k})\V{B}^*_3(\V{k}) -\V{B}_2^*(\V{k})\V{B}_3(\V{k})]}{\omega_{SC}/V_{SW}}\\
&\times \delta[\omega_{SC}-
(\V{k}\cdot \V{V_{SW}} + \omega)],
\end{split}
\end{equation}
where $\V{V}_{SW}$ is the solar wind velocity.  Normalizing $H_m
^{r'}$ by the trace power $|\V{B}(k_1)|^2$ yields the \emph{normalized reduced
fluctuating magnetic helicity}, $\sigma_m'=k_1 H_m
^{r'}(k_1)/|\V{B}(k_1)|^2$, where $k_1\equiv
\omega_{SC}/V_{SW}$ is the projection of the wavevector along the 
direction of the solar wind.  The normalized reduced fluctuating
magnetic helicity is bound between $-1$ and $1$.  For simplicity,
throughout this paper the simplified term ``magnetic helicity" is used
instead of the more cumbersome ``normalized reduced fluctuating magnetic
helicity".

Spacecraft measurements of $\sigma_m'$ in the near-Earth solar wind at
low spacecraft-frame frequencies, 
$f \ll 1 \mbox{ Hz}$, typically give values that fluctuate
about zero, which was originally interpreted as an admixture of waves
with left-handed $(\sigma_m' \simeq -1)$ and right-handed $(\sigma_m'
\simeq +1)$ magnetic helicities \citep{Matthaeus:1982a}.  Based on 
the eigenfunctions of the linear Vlasov-Maxwell dispersion relation,
\cite{Gary:1986} later showed that, at large-scales $k \rho_i \ll 1$,
the linear waves have very small intrinsic magnetic helicity,
$\sigma_m' \simeq 0$, which eliminated the need for an explanation
in terms of a mixture of waves with left- and right-handed helicities.
At higher spacecraft-frame frequencies, $f \sim 1 \mbox{ Hz}$,
\emph{in situ} measurements of the magnetic field fluctuations produced an
identifiably nonzero value of $\sigma_m'$
\citep{Goldstein:1994,Leamon:1998a}. This was initially interpreted as 
evidence of the damping of left-hand polarized ion cyclotron waves (ICWs) and the
persistence of right-hand polarized whistler waves.  

Subsequently, it was shown that an anisotropic turbulent spectrum of
\Alfven waves and kinetic \Alfven waves (KAWs) naturally reproduces
both the low and high frequency measurements \citep{Howes:2010a}. As
KAWs have preferentially perpendicular wavevectors, $k_\perp >
k_\parallel$, and ICWs and whistler waves require sufficiently large
parallel wavevectors, $k_\parallel d_i \gtrsim 1$,
where $\perp$ and $\parallel$
refer to orientation with respect to the \emph{local} mean magnetic
field $\V{B_0}$, the KAW model has the added benefit of being
compatible with the predominately perpendicular cascade of energy
expected from anisotropic magnetized turbulence theories
\citep{Goldreich:1995,Boldyrev:2006}.  Two recent studies
\citep{He:2011a,Podesta:2011a}, analyzed the magnetic helicity
of solar wind fluctuations as a function of the wave period
$T$ in the spacecraft frame and the angle $\theta$ between 
the solar wind velocity $\V{V_{SW}}$ and
$\V{B_0}$, and discovered two distinct signatures
at angles perpendicular and parallel to $\V{B_0}$ at spacecraft-frame 
frequencies $f \sim 1$ Hz. These observations have been interpreted as
two separate populations of fluctuations 
at kinetic scales with wavevectors oriented parallel and perpendicular
to the local mean magnetic field \citep{Podesta:2011b,He:2012}.


We here explore the properties of a model for solar wind turbulence
that reproduces the observed behavior of the magnetic helicity.  In
particular, we show that the properties of the model
are tightly constrained by spacecraft observations of
the magnetic helicity plotted as a function of period and angle.  This
study employs the \emph{synthetic spacecraft data method}
\citep{Klein:2012}, in which synthetic time series measurements 
of the magnetic field---measurements that may be directly compared to
spacecraft observations---are generated by sampling along a trajectory
through the model plasma volume.  The turbulent fluctuations in this
model are derived from a distribution of linear wave modes with random
phases, where the particular distribution of wave power is guided by
modern plasma turbulence theory, and where the physical properties of
the wave modes are derived from linear kinetic plasma physics. The
model's ability to reproduce the observed magnetic helicity structure
is used to constrain the underlying parameters describing the
distribution of turbulent wave power, shedding light on the nature of
the turbulent fluctuations in the weakly collisional solar wind
plasma.
 
The remainder of the paper is organized as follows.
Section~\ref{sec:mag_hel} provides background on the germane aspects
of recent \textit{in situ} magnetic helicity measurements as well as a
discussion of the underlying linear kinetic plasma theory.  A
description of and justification for the underlying turbulence model
is given in Section \ref{sec:turbulence}.  The general method used to
generate the synthetic data and the particular results are found in
Section \ref{sec:Synth}. Discussion of these results and a conclusion
follow in Sections
\ref{sec:discussion} and  \ref{sec:summary} respectively.



\section{Magnetic Helicity}
\label{sec:mag_hel}

In this section, we review the observations of magnetic helicity in
solar wind turbulence and outline the linear kinetic physics that
determines the magnetic helicity of each of the characteristic linear
wave modes.


\subsection{Spacecraft Measurements}
\label{sec:inSitu}



\begin{figure}[t]
\begin{center}
\includegraphics[width=7.75cm,viewport=0 0 530 420, clip=true]
{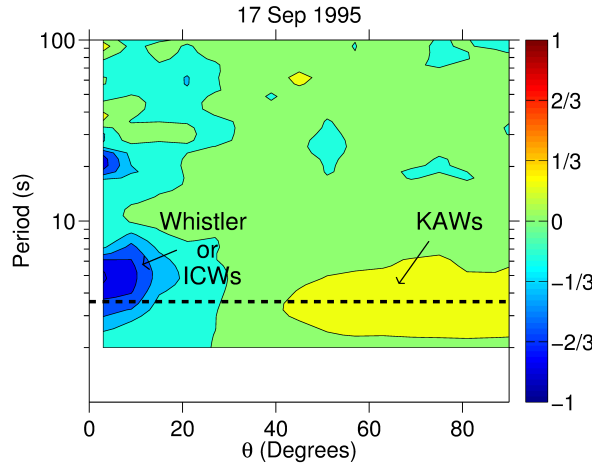}
\caption{Magnetic helicity $\sigma_m'$ computed from observations 
by the \emph{Ulysses} spacecraft, plotted as a function of period and
angle for interval I.  Previous studies have suggested that the
perpendicular signature is due to kinetic
\Alfven waves (KAWs) and the parallel signature is due to either
whistler or ion cyclotron waves (ICWs)
\citep{Podesta:2011a,He:2011a}.
The horizontal dashed line indicates the period
approximately corresponding to $k \rho_i=1$.}
\label{fig:sept17}
\end{center}
\end{figure}

Although magnetic helicity has long been used to investigate the 
nature of turbulent fluctuations in the solar wind 
\citep{Matthaeus:1982a,Matthaeus:1982b,Gary:1986,Leamon:1998a,
Goldstein:1994,Hamilton:2008,Howes:2010a}, only recently has it been
shown by plotting the magnetic helicity of the turbulent fluctuations
as a function of period and angle that there exist two distinct
signatures of the magnetic helicity at scales near $k \rho_i=1$
\citep{He:2011a,Podesta:2011a}. Employing a wavelet analysis of
\textit{in situ} magnetic field measurements, both studies calculated
$\sigma_m'$ as a function of the wave period $T$ in the spacecraft
frame and the angle
$\theta$ between the solar wind velocity $\V{V}_{SW}$ and
\emph{local} mean magnetic field $\V{B}_0$. This method showed that,
in a majority of the measurements, there appear two distinct
signatures: (i) a broad feature with positive magnetic helicity at
large angles centered around $\theta \sim 90^\circ$, and (ii) a
more narrow feature with negative magnetic helicity at small angles
$\theta \sim 0^\circ$. We denote these distinct features as the
\emph{perpendicular} and \emph{parallel signature}, respectively.

These features are clearly visible in Figure~\ref{fig:sept17}, a plot
of the magnetic helicity $\sigma_m'$ as a function of period $T$ and
angle $\theta$, computed from one of the same \emph{Ulysses}
spacecraft data intervals as analyzed by
\cite{Podesta:2011a}.  We note that if the measurements are summed over all
angles $\theta$ to obtain a frequency spectrum of the magnetic
helicity $\sigma_m'(\omega_{SC})$, the signature of the negative
magnetic helicity at small angles $\theta \sim 0^\circ$ is
completely masked by the positive helicity at large angles
$\theta \sim 90^\circ$ \citep{Podesta:2011a}.

An important difference in the results of \cite{He:2011a} and \cite{Podesta:2011a}
is the magnitude of the maximum value of the parallel magnetic helicity signature. 
In \cite{He:2011a}, the observed maximum value of $|\sigma_m|$ for the parallel signature 
is approximately $0.4$
while \cite{Podesta:2011a} found a maximum value that is consistently near
$1.0$. This difference persists even when analyzing essentially the same 
interval of STEREO data.
The different results may be due to subtle differences between the analysis
techniques employed in the two studies; further work is needed to 
identify the cause of these differences. All analysis undertaken and results presented
in this study are based upon the observations found in \cite{Podesta:2011a}.

Several hypotheses have been proposed to account for the qualitative
features of the magnetic helicity plot seen in
Figure~\ref{fig:sept17}.  The perpendicular signature was suggested to
arise from an anisotropic distribution of KAWs with $k_\perp \gg
k_\parallel$ \citep{He:2011a,Podesta:2011a} or from a spectrum of
whistler waves \citep{He:2011a}. The parallel signature was proposed
to arise from ICWs propagating along the magnetic field away from the sun
\citep{He:2011a,Podesta:2011a} or whistler waves propagating along the
magnetic field toward to the sun
\citep{Podesta:2011a,Podesta:2011b}. Temperature anisotropy
instabilities, such as the electromagnetic ion cyclotron instability
and the parallel firehose instability, have been proposed to generate
these parallel propagating waves
\citep{Podesta:2011a,Podesta:2011b}. The maximum growth rate of these
instabilities occurs at length scales near the ion inertial length,
$k_\parallel d_i \sim 1$ \citep{Podesta:2011b}, as shown in
Figure~\ref{fig:unstable}, which may explain why the parallel
signature is localized near periods $T \sim 5\mbox{ s}$.

\begin{table*}[t]
\begin{center}
\begin{tabular}{|c|l|c|c|c|c|}
\hline
Interval & Date & $\beta_i$ & $\rho_i/V_{SW} \ (s)$ & R (AU)& \\
\hline
\textbf{I}   & Sept. 17 14:38 to Sept 18 4:46 1995 & $1.7$ & $0.544$&
$2.36$& $1^{st}$ Northern Pass\\
\textbf{II}  & May 23 00:32 to 22:20 1995          & $1.0$ & $0.312$&
$2.36$&$1^{st}$ Northern Pass\\
\textbf{III} & July 6 22:12 to July 7 05:51 1994.  & $1.5$ & $0.662$&
$2.76$&$1^{st}$ Southern Pass\\
\hline
\end{tabular}
\caption{Selected \emph{Ulysses} data intervals from \cite{Podesta:2011a}
chosen for direct comparison to synthetic spacecraft data models.
}
\label{tab:intervals}
\end{center}
\end{table*}

In this study, we construct the magnetic helicity plot for three
intervals of \emph{Ulysses} data specified in
Table~\ref{tab:intervals}---intervals that have been previously
analyzed in \cite{Podesta:2011a}---for direct comparison to magnetic
helicity plots generated by the synthetic spacecraft data method
\citep{Klein:2012}.  A detailed discussion of the \textit{in situ}
data analysis techniques can be found in \cite{Podesta:2011a}.  The
first two intervals are from the first northern pass of \emph{Ulysses}
while the third is from the southern pass. Contours of $\sigma_m'$
from these data sets are presented in the left hand column of
Figure~\ref{fig:grid}.  The ion plasma beta, 
$\beta_i \equiv 8 \pi n_i T_i/B_0^2$, is 
$\beta_i \ge 1$ for these three periods, 
where $T_i$ is the
ion temperature in energy units and $n_i$ is the ion density. No
interval from the \cite{Podesta:2011a} data set had $\beta_i <
1$. Intervals \textbf{I} and \textbf{III} both have the parallel and
perpendicular signatures as previously described, while interval
\textbf{II} was chosen as an abnormal period with no visible
perpendicular structure. The key question that we ask is, \emph{what
do these magnetic helicity measurements tell us about the nature of
the turbulent fluctuations in the solar wind?}  By making detailed
quantitative comparisons to magnetic helicity plots generated by the
synthetic spacecraft data method, we hope to illuminate the
characteristics of the underlying turbulent fluctuations in the solar wind.


\subsection{Linear Kinetic Physics}
\label{sec:linear}

The kinetic physics of the characteristic plasma waves in a weakly
collisional plasma such as the solar wind is used to guide our
intuition about the magnetic helicity of turbulent fluctuations.  To
investigate the linear wave properties in a weakly collisional plasma,
we employ the linear Vlasov-Maxwell dispersion relation for a fully
ionized proton and electron plasma with isotropic Maxwellian
equilibrium distribution functions.  The linear Vlasov-Maxwell
dispersion relation can be expressed as a function of five
dimensionless parameters: $k_\perp
\rho_i$, $k_\parallel d_i$, $\beta_i$, $T_i/T_e$, and $v_{t_i}/c$.
Here, $\rho_i =v_{t_i}/\Omega_i$ is the ion Larmour radius,
$d_i=c/\omega_{pi}=\rho_i/\sqrt\beta_i$ is the ion inertial length,
$v_{t_i}= \sqrt{2T_i/m_i}$ is the ion thermal speed (where Boltzmann's
constant has been absorbed to express temperature in units of energy),
$\Omega_i=e B_0/ m_i c$  is the ion cyclotron frequency 
$\omega_{pi}=\sqrt{4 \pi n_i e^2/m_i}$ is the ion
plasma frequency , and $c$ is the speed of light.  

The three familiar
wave modes in magnetohydrodynamics (MHD)---the \Alfven wave and the
fast and slow magnetosonic waves---have kinetic counterparts in a
collisionless plasma that, at large scales $k \rho_i \ll 1$, have
properties similar to the MHD modes \citep{Klein:2012}.  Each of these
non-dispersive MHD-like waves transitions to a related dispersive wave
mode (\emph{e.g.}, kinetic \Alfven waves or whistler waves) at smaller
scales corresponding to $k_\perp \rho_i \gtrsim 1$ or $k_\parallel d_i
\gtrsim 1$. It has been shown that the compressible energy in the
inertial range of turbulence in the solar wind, which constitutes 10\%
or less of the total turbulent energy at these scales, consists of
kinetic slow waves, with a negligible contribution from kinetic fast
waves
\citep{Howes:2012a,Klein:2012}. 
Although this finding merits further
investigation of the magnetic helicity of the kinetic slow waves, 
in this study we focus on the physics of the kinetic fast waves and
\Alfven waves and their dispersive extensions at small scales.

\begin{figure}[t]
\begin{center}
\includegraphics[width=8cm,viewport= 5 20 137 205, clip=true]
{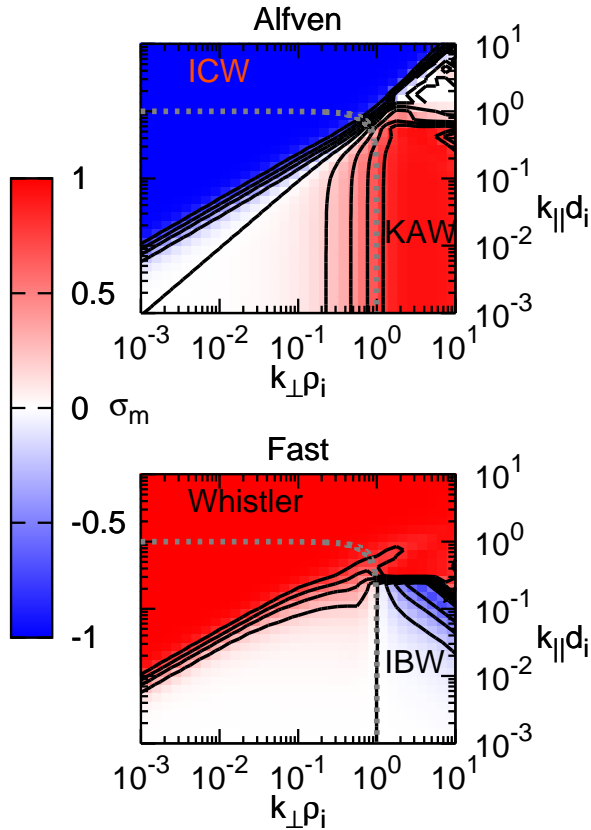}
\caption{Log-log contour maps of $\sigma_m(\V{k})$ for the
\Alfven (top) and fast (bottom) modes as functions of $k_\perp \rho_i$ and
$ k_\parallel d_i $. The regions of interest for
the ICW, KAW, and whistler waves are indicated. The dashed gray line
demarcates $k\rho_i=1$, the transition from non-dispersive to
dispersive wave behavior.}

\label{fig:sigma_k}
\end{center}
\end{figure}

The \Alfven wave transitions to the ion cyclotron wave (ICW) in the
limit of small parallel scales $k_\parallel d_i \gtrsim 1$ and to the
kinetic \Alfven wave (KAW) in the anisotropic limit of small
perpendicular scales $k_\perp \rho_i \gtrsim 1$ and $k_\perp \gg
k_\parallel$.  The kinetic fast wave transitions to the whistler wave
in the limit of small parallel scales $k_\parallel d_i \gtrsim 1$ and
to the ion Bernstein wave in the limit of small perpendicular scales
$k_\perp \rho_i \gtrsim 1$ and $k_\perp \gg k_\parallel$
\citep{Howes:2009,Verdon:2009}.
In this study we focus on the magnetic helicity
signatures of the kinetic \Alfven wave, ion cyclotron wave, and
whistler wave. The regions in the $(k_\perp,k_\parallel)$ plane where
these wave modes exist are depicted in Figures~\ref{fig:sigma_k}
and~\ref{fig:region}.

For an individual plane wave with wavevector $\V{k}$, we employ
the linear Vlasov-Maxwell dispersion relation to determine the
eigenfunction of the kinetic fast wave and \Alfven wave and compute
the normalized fluctuating magnetic helicity $\sigma_m(\V{k}) \equiv k
H'_m(\V{k}) / | \V{B}(\V{k})|^2$ as a function of $k_\perp \rho_i$ and
$k_\parallel d_i$, where $H'_m(\V{k})$ is defined in \cite{Howes:2010a}.  
In Figure~\ref{fig:sigma_k}, we plot
$\sigma_m(\V{k})$ for the kinetic fast and \Alfven modes for plasma
parameters $\beta_i=1$, $T_i/T_e=1$, and $v_{t_i}/c=10^{-4}$. We adopt
the convention that $\omega>0$, so that the wave propagation direction
along the magnetic field is determined by the sign of $k_\parallel$.
The plot in Figure~\ref{fig:sigma_k} employs $k_\parallel>0$. 

In Figure~\ref{fig:sigma_k}, we show that large scale, isotropic
fluctuations of both the kinetic fast (bottom) and \Alfven (top) modes
have a magnetic helicity of nearly zero. For nearly parallel
wavevectors $k_\parallel \gg k_\perp$ (the upper left corner of each
of the plots in Figure~\ref{fig:sigma_k}), the kinetic fast and
whistler modes have a right-handed (positive) magnetic helicity
$\sigma_m\simeq 1$, while the \Alfven and ICW modes have a left-handed
(negative) magnetic helicity $\sigma_m\simeq -1$. For nearly
perpendicular wavevectors with $k_\perp \gg k_\parallel$ and $k_\perp
\rho_i \gtrsim 1$, the kinetic fast and ion Bernstein modes maintain
zero magnetic helicity $\sigma_m\simeq 0$, while the KAW mode develops
a right-handed (positive) magnetic helicity $\sigma_m\simeq 1$.  Note
that the magnetic helicity $\sigma_m$ has odd parity with respect to
$k_\parallel$, so that modes traveling in the opposite direction
with $k_\parallel <0$ will have the opposite signs of the magnetic
helicity $\sigma_m$. Therefore, in the case of parallel propagating
modes, it is not possible using magnetic helicity alone to distinguish
between whistler waves traveling up the mean magnetic field and ICWs
traveling down the mean magnetic field. Suggestions to break this
degeneracy using other physical characteristics are discussed in
Section~\ref{sec:discussion}.
Linear eigenfunction calculations for lightly damped kinetic slow waves 
with $k_\parallel d_i \ll 1$ (not shown) find $\sigma_m' \sim 0$,
suggesting these fluctuations would not significantly contribute 
to \emph{in situ} observations of magnetic helicity. 

The relevance of the magnetic helicity of the linear kinetic
modes to the magnetic helicity of turbulent fluctuations in the solar
wind is based on the premise that, to lowest order, some properties of
the turbulent fluctuations are governed by the linear response of the
plasma. This concept is central to the \emph{quasilinear premise}
\citep{Klein:2012}, the idea that some properties of magnetized plasma
turbulence can be understood by modeling the turbulence as a
collection of randomly phased, linear waves. In this picture, the
nonlinear turbulent interactions serve to transfer energy from one
linear wave mode to another---thus, the picture is quasilinear.
Detailed arguments in support of the quasilinear premise are presented
in \citet{Klein:2012}.

Adopting the quasilinear premise, the amplitude and phase relationships among different 
components of a single plane-wave or mode, e.g., the three components of the magnetic 
field, are governed by the linear eigenfunctions of the characteristic plasma mode. The 
phase correlations among different plane-wave modes dictated by the nonlinear 
interactions, correlations that are responsible for the appearance of coherent structures 
and intermittency, are not described by our model.

Modern MHD turbulence theory suggests that a state of
critical balance is maintained in strong plasma turbulence
\citep{Higdon:1984,Goldreich:1995,Boldyrev:2006}, implying that the
linear terms responsible for the amplitude and phase relationships
within a single plane-wave mode contribute to the evolution at the
same order as the nonlinear terms responsible for the amplitude and
phase relationships among different plane-wave modes. Therefore, a
direct comparison of the predictions of linear wave properties
of kinetic plasma physics through the synthetic spacecraft data
method to \emph{in situ} solar wind measurements can provide insight
into the nature of the turbulent fluctuations present in the solar
wind. 

We do not directly address the possible mechanisms through which coherent
structures or other nonlinear structures \citep{Osman:2012,Perri:2012,Roberts:2013} 
might produce the observed
parallel and perpendicular signatures in solar wind measurements of magnetic helicity.
Rather, we show that linear fluctuations arising through plausible
physical mechanisms are sufficient to produce the observed
magnetic helicity signatures. Further work is necessary to establish 
a theory through which coherent structures and other nonlinear mechanisms 
could replicate these magnetic helicity plots.

In this paper, we adopt the quasilinear premise and employ the
synthetic spacecraft data method to generate predicted plots of the
magnetic helicity $\sigma_m'$ as a function of period $T$ and angle
$\theta$.  A detailed quantitative comparison of these synthetic
magnetic helicity plots to observations yields stringent new constraints
on the nature of turbulence in the solar wind.


\section{Physically Motivated Turbulence Model}
\label{sec:turbulence}


The magnetic helicity observations \citep{He:2011a,Podesta:2011a} that
show distinct perpendicular and parallel signatures, as described in
Section~\ref{sec:inSitu} and exemplified by Figure~\ref{fig:sept17},
provide an important probe of the turbulence in the solar wind.
Qualitatively, the perpendicular magnetic helicity signature is
believed to arise from an anisotropic cascade of Alfv\'enic
fluctuations that mediates the nonlinear transfer of energy from large
to small scales \citep{Howes:2010a,He:2011a,Podesta:2011a}, and the
parallel signature is believed to arise from fluctuations driven at
ion kinetic length scales by temperature anisotropy instabilities in
the spherically expanding solar wind flow
\citep{Hellinger:2006,Bale:2009,Podesta:2011a,Podesta:2011b}.
In this section, we present the scientific reasoning supporting this
qualitative interpretation of the physics underlying the two observed
magnetic helicity signatures, and then we describe a quantitative
model of the turbulent fluctuations in the solar wind that we use to
create synthetic magnetic helicity data for comparison to the
observations.


\subsection{Scientific Background for Turbulence Model}
\label{sec:justification}


The synthetic spacecraft data method is useful in the investigation of
solar wind turbulence only if a realistic model of the turbulent
fluctuations can be constructed.  To accomplish this goal, we exploit
the significant advances in the understanding of MHD and kinetic
turbulence made in the last two decades and the resulting
theory of anisotropic \Alfvenic turbulence
\citep{Goldreich:1995,Quataert:1998,Boldyrev:2006,Howes:2008c,Schekochihin:2009}.

Accounting for the anisotropic nature of the dynamics in a magnetized
plasma, \citet{Goldreich:1995} conjectured that strong incompressible
MHD turbulence would maintain a state of critical balance, implying a
preferential nonlinear transfer of energy to small perpendicular
scales. This anisotropic cascade leads to small-scale fluctuations
that are highly elongated along the direction of the mean magnetic
field, a consequence of the wavenumber anisotropy $k_\perp\gg
k_\parallel$, a condition supported by numerical simulations
\citep{Cho:2000b,Maron:2001,Cho:2003a,TenBarge:2012a} and observations
of turbulence in the solar wind
\citep{Horbury:2008,Podesta:2009,Chen:2010a,
Sahraoui:2010,Narita:2011,Roberts:2013,Podesta:2013}. 
We refer to this anisotropic energy distribution in wavevector space as a
\emph{critically balanced distribution}. 

Compressible MHD turbulence simulations demonstrate an isotropic
cascade of fast waves and critically balanced distributions of \Alfven
and slow waves \citep{Cho:2003a}. Recent results using the synthetic
spacecraft data method to explore the correlation between density and
parallel magnetic field fluctuations have demonstrated that negligible
energy exists in fast wave fluctuations in the inertial range of solar
wind turbulence
\citep{Howes:2012a,Klein:2012}, so we exclude fast wave fluctuations
from the inertial range of our turbulence model. 

When the \Alfvenic turbulent cascade reaches the perpendicular scale
of the ion (proton) Larmor radius, $k_\perp \rho_i
\sim 1$, it has been proposed that the turbulence transitions from a
cascade of MHD
\Alfven waves to a cascade of kinetic \Alfven waves
\citep{Leamon:1998a,Leamon:1999,Howes:2008a,Howes:2008c,Schekochihin:2009,Howes:2011b},
a hypothesis supported by kinetic numerical simulations of plasma
turbulence
\citep{Howes:2008b,Howes:2011a,TenBarge:2012a,
TenBarge:2013a,TenBarge:2013c}
and by observations of solar wind turbulence at dissipation range scales
\citep{Sahraoui:2010,Salem:2012,Podesta:2013,Chen:2013a}. Based on this
wealth of theoretical, numerical, and observational evidence, the
major component of the turbulent fluctuations in our model is an
anisotropic \Alfvenic turbulent cascade, consisting of a critically
balanced distribution of \Alfven waves at scales $k_\perp \rho_i \ll
1$, transitioning to a critically balanced distribution of kinetic
\Alfven waves at scales $k_\perp \rho_i \gtrsim 1$.

Observational studies have demonstrated that the temperature
anisotropy, $T_\perp/T_\parallel$, of ion (proton) velocity
distributions in the weakly collisional solar wind is bounded by the
marginal stability criteria of kinetic temperature anisotropy
instabilities
\citep{Kasper:2002,Hellinger:2006,Bale:2009}. In the spherically expanding
solar wind flow, evolution of the temperature anisotropy
can drive the plasma into an unstable regime,
exciting fast-growing kinetic instabilities that tap the free energy
in the anisotropic velocity distributions to drive fluctuations at
characteristic kinetic scales in the plasma. We hypothesize here that
\emph{these kinetic-instability-driven waves may persist in the solar wind
alongside the \Alfvenic turbulent fluctuations that mediate the
nonlinear transfer of energy from large to small scales.}

Specifically, for an ion temperature anisotropy $T_{\perp
i}/T_{\parallel i}>1$, the electromagnetic ion cyclotron (EMIC)
instability may generate ion cyclotron waves with a rapid growth rate,
$\gamma/\Omega_i \sim 1$, that peaks for parallel wavenumbers near the
proton inertial length, $k_\parallel d_i \sim 1$.  For a proton
temperature anisotropy $T_{\perp i}/T_{\parallel i}<1$, the parallel
firehose instability may generate whistler waves, also with a positive
growth rate, $\gamma/\Omega_i \sim 0.1$, that peaks for parallel
wavenumbers near the proton inertial length, $k_\parallel d_i \sim 1$.
Recall that the ion cyclotron wave is the $k_\parallel d_i \gtrsim 1$
limit of the \Alfven wave solution to the linear Vlasov-Maxwell
dispersion relation, and the whistler wave is the $k_\parallel d_i \gtrsim
1$ limit of the fast magnetosonic wave solution. 

\begin{figure}[t]
\begin{center}
\includegraphics[width=8.5cm,viewport=20 25 175 155, clip=true]
{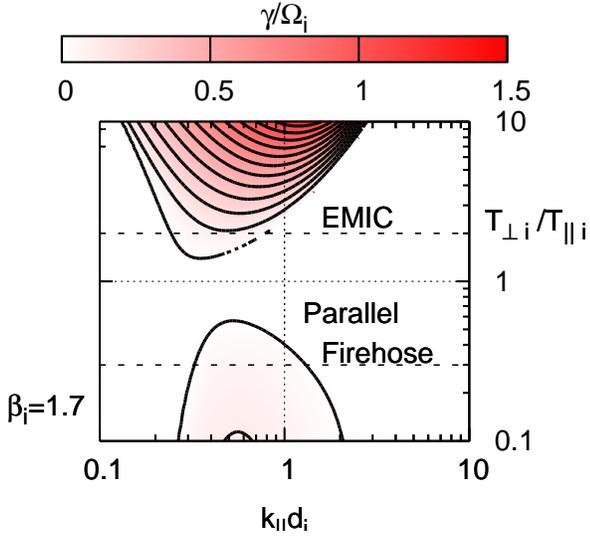}
\caption{Contour map of positive growth rates $\gamma/\Omega_i$
  parametrized by the ion temperature anisotropy
  $T_{\perp i}/T_{\parallel i}$
  and the parallel scale size
  $k_\parallel d_i$ for the electromagnetic ion cyclotron (EMIC) and
  parallel firehose instabilities from linear kinetic theory with
  $k_\perp =0$.  We take $\beta_i=1.7$ to match conditions for
  interval \textbf{I}; the resulting $\gamma/\Omega_i$ is similar for
  the parameters of the other two intervals, though the parallel
  firehose instability disappears for $\beta_i <1$. The dashed
  horizontal lines represent approximate bounds on
  $T_{\perp,i}/T_{\parallel,i}$ in the solar wind near 1 AU for
  $\beta_i \gtrsim 1$ \citep{Hellinger:2006,Bale:2009}.  No minority
  ions effects are included in this calculation, which may drastically
  affect the damping rates \citep{Podesta:2011b}.}
\label{fig:unstable}
\end{center}
\end{figure}

For plasma parameters relevant to interval \textbf{I} and assuming
that unstable modes have $k_\perp =0$, a contour plot of the instability
growth rates for the EMIC and parallel firehose instabilities as a
function of $k_\parallel d_i$ and $T_{\perp i}/T_{\parallel i}$ is
presented in Figure~\ref{fig:unstable}, which was generated using
equations (11-2) and (11-3) from \cite{Stix:1992}. This plot
shows that the instability growth rates are highly localized in scale
to $k_\parallel d_i \sim 1$, suggesting that these fluctuations are
generated locally, and are likely to be unrelated to the \Alfvenic
turbulent fluctuations responsible for the cascade of energy from
large scales.  The growth rates do not drastically differ for
intervals \textbf{II} and \textbf{III}, though the parallel firehose
instability disappears for $\beta_i <1$. Therefore, the secondary
component of the fluctuations measured in the solar wind consists of a
distribution of either ion cyclotron waves (ICWs) or whistler waves
with wavevectors nearly parallel to the local mean magnetic field and
$k_\parallel d_i \sim 1$.

As shown in Section~\ref{sec:linear}, the normalized fluctuating
magnetic helicity $\sigma_m(\V{k})$ for parallel waves is
$\sigma_m(\V{k})=+1$ for whistler waves and $\sigma_m(\V{k})=-1$
for ion cyclotron waves.  Therefore, the parallel signature observed
in spacecraft measurements with magnetic helicity $\sigma_m'\simeq
-1$, as shown in Figure~\ref{fig:sept17}, can be explained by either
ICWs propagating unidirectionally away from the sun or by whistler
waves propagating unidirectionally toward the sun
\citep{Podesta:2011b}. By accounting for a relative drift velocity 
along the magnetic field between alpha particles and protons, a
characteristic frequently observed in the
weakly collisional solar wind, \cite{Podesta:2011b} showed that an
asymmetry may develop in the generation of unstable waves, with the
result that anti-sunward ICWs are preferentially driven by the EMIC
instability and that sunward whistler waves are preferentially driven
by the parallel firehose instability and that the
presence of an alpha particle drift can lead to an enhancement
of the growth rates for both of these instabilities.
Either of these instabilities
would then be able to account for the observed parallel magnetic
helicity signature. In addition, it has been shown that accounting for
the temperature anisotropy of the alpha particles 
\citep{Verscharen:2013b} and the presence of 
minor ions  \citep{Matteini:2012} can significantly alter 
the growth rates of kinetic temperature anisotropy instabilities.

It should also be noted that an alternative interpretation of the
parallel magnetic helicity signature is that it arises as part of the
turbulent cascade \citep{He:2011a,He:2012}. This viewpoint conflicts
with the strong theoretical and numerical evidence discussed above
that supports an anisotropic cascade with energy preferentially
transferred to small perpendicular scales and little energy
transferred to small parallel scales. Rather than the energy in the
fluctuations responsible for the parallel magnetic helicity signature
coming from nonlinear energy transfer from large scale fluctuations
through a turbulent cascade, we adhere to the viewpoint that the
energy is directly injected into these parallel fluctuations at a
scale $k_\parallel d_i \sim 1$ by kinetic instabilities 
\citep{Podesta:2011a,Podesta:2011b}.  Therefore, the distribution 
of power in parallel modes in our turbulence model is restricted to
the nearly parallel wavevectors at which the instability growth rates
peak, as seen in Figure~\ref{fig:unstable}.


\subsection{Turbulence Model Parameters}
\label{sec:parameters}


\begin{figure}[t]
\begin{center}
\includegraphics[width=10.0cm,viewport=25 0 200 125, clip=true]
{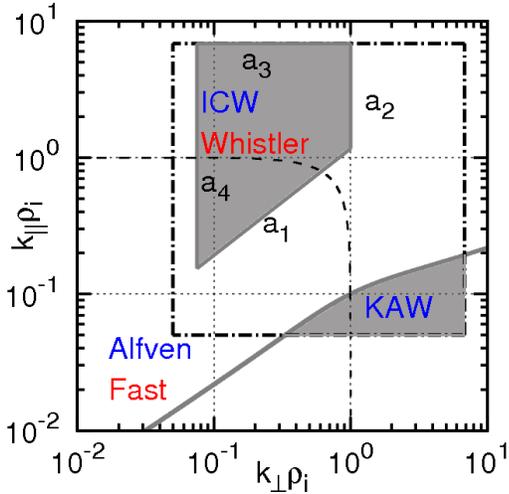}
\caption{Regions of  $(k_\perp \rho_i,k_\parallel \rho_i)$ parameter space in 
which different wave modes exist that are associated with the \Alfven
(blue) and fast (red) wave solutions.  The wavevector region used in
creating the synthetic data is indicated by the dot-dashed box, while
the limits of the perpendicular and parallel populations are indicated
with solid grey lines. The perpendicular region is bounded above by
the critical balance line while the parallel region is defined by four
parameters $a_1,a_2,a_3$ and $a_4$. In this study, $a_1$ and $a_2$ are
kept constant and the variation of $a_3$ and $a_4$ fix the maximum and
minimum values of $k_\parallel \rho_i$.}
\label{fig:region}
\end{center}
\end{figure}

As discussed above, our model for the turbulent fluctuations in the
solar wind contains two main components: (i) a critically balanced
distribution of \Alfven and kinetic \Alfven waves that is responsible
for the perpendicular magnetic helicity signature; and (ii) a
distribution of either ion cyclotron waves (ICWs) or whistler waves
with nearly parallel wavevectors peaked around a magnitude
$k_\parallel d_i \sim 1$ that is responsible for the parallel magnetic
helicity signature. A schematic diagram of the regions of wavevector
space $(k_\perp,k_\parallel)$ occupied by these two components is
shown in Figure~\ref{fig:region}. 
This figure shows a logarithmic
representation of the component of the wavevector perpendicular to the
local mean magnetic field,
$k_\perp \rho_i$, versus the parallel component of the wavevector
$k_\parallel \rho_i$. Unlike in previous sections, we have chosen
to normalize both the parallel and perpendicular wavevectors
by the same scale length to insure a uniformly spaced grid for
the creation of our synthetic spacecraft data. Wave power
is assumed to be distributed axisymmetrically about the magnetic field
direction in the perpendicular plane, so this two-dimensional diagram
is sufficient to represent the three-dimensional distribution of power
used to create our synthetic spacecraft data. 

The dot-dashed box in Figure~\ref{fig:region} indicates the
perpendicular and parallel size of our synthetic plasma volume as well
as the perpendicular and parallel resolution. The lower gray region
(KAW) represents the portion of the critically balanced cascade of
\Alfven and kinetic \Alfven waves that can be represented within the
domain size and resolution limits of our synthetic plasma volume.  The
upper gray region (ICW or Whistler) represents the distribution of
parallel wave modes populating our synthetic plasma volume. Modes
with $k_\perp =0$ or $k_\parallel =0$ are both excluded from the synthetic volume.
In addition to parameters that specify precisely the boundaries of these
regions of wavevector space, our turbulence model also requires (i)
two parameters that specify the imbalance of wave energy fluxes in the
sunward versus anti-sunward directions for the parallel and
perpendicular wave populations and (ii) one parameter that specifies
the relative power between the perpendicular and parallel wave
distributions. Below we define in detail each of these parameters for
our turbulence model.

Power for the perpendicular distribution of \Alfven and kinetic \Alfven
waves is only found in
the wavevector region below the line defining critical balance 
\citep{Goldreich:1995,Howes:2008b,Howes:2011b,TenBarge:2012a},
\begin{equation}
k_\parallel\rho_i =(k_0\rho_i)^{1/3}\left[ \frac{(k_\perp
\rho_i)^{2/3} + (k_\perp\rho_i)^{7/3}} {1+(k_\perp\rho_i)^2}\right],
\end{equation}
shown by the gray line in Figure~\ref{fig:region}.  This curve
has a single parameter, the \emph{isotropic driving wavenumber}, $k_0
\rho_i$ \citep{Howes:2008b}, which is physically determined by the outer scale of the 
inertial range observed in the solar wind (at the low frequencies
where the measured magnetic energy frequency spectrum steepens from a
spectral index of $-1$ to $-5/3$). For this study, we take the
isotropic scale $k_0 \rho_i= 10^{-3}$, a value within observationally
suggested limits \citep{Bale:2005,Schekochihin:2009}. It is believed
that this anisotropic distribution of \Alfvenic fluctuations dominates
the turbulent power in the solar wind, so the energy as a function of
wavenumber magnitude $k\rho_i$ (assuming the Taylor hypothesis
[\cite{Taylor:1938}] to convert from spacecraft-frame frequency to
wavenumber) is well constrained by the magnetic energy spectrum
observed in the solar wind. Therefore, \emph{in situ} measurements of
the inertial and dissipation ranges of solar wind turbulence
\citep{Sahraoui:2009,Alexandrova:2009,Kiyani:2009,Sahraoui:2010,
Chen:2010b,Roberts:2013,Podesta:2013}
lead us to specify a turbulent spectrum which scales as $k^{-5/3}$ at
low wavenumbers and breaks to $k^{-2.8}$ at $k \rho_i \sim 1$. Note
that using a $k^{-3/2}$ spectrum \citep{Boldyrev:2006} instead of
$k^{-5/3}$ in the inertial range has no qualitative impact on the
resulting synthetic magnetic helicity measurements.

Note that the intersection of the critically balanced distribution
with the parallel domain size of the synthetic plasma volume (the
lower edge of the dot-dashed box in Figure~\ref{fig:region}) cuts off
the representation of \Alfvenic wave modes at a minimum $k_\perp\rho_i
\simeq 0.3$. This lack of large scale power in the \Alfvenic turbulent
cascade prevents us from accurately describing the magnetic helicity
signature at long periods. But, as shown in Figure~\ref{fig:sept17},
the dominant features of the magnetic helicity plot occur at shorter
periods $T \le 10$~s; these shorter period fluctuations are associated
with the dynamics at kinetic scales $k \rho_i \sim 1$, where the
intrinsic magnetic helicity $\sigma_m(\V{k})$ is nonzero, as shown in
Figure~\ref{fig:sigma_k}.  Therefore, this limitation does not prevent
this study from constraining the nature of the turbulent fluctuations
at kinetic scales using the magnetic helicity as a probe.

The parallel component of the turbulence model consists of either
ICWs or whistler waves distributed in the region of wavevector
space with nearly parallel wavevectors given by the upper gray region
in Figure~\ref{fig:region}.  The boundaries of this region are
specified by four parameters: $a_1$, $a_2$, $a_3$, and $a_4$. The
limits of the region in the perpendicular direction are given by
$(k_\perp
\rho_i)_{{min}}=a_4$ and $(k_\perp \rho_i)_{{max}}=a_2$, and
the maximum in the parallel direction is given by $(k_\parallel
\rho_i)_{{max}}=a_3$. 

The lower boundary of the parallel region, associated with
$a_1$, is chosen to coincide with the slope of the $\sigma_m(\V{k})=-0.8$ contour in
Figure~\ref{fig:sigma_k}; if the boundary is moved further down
to smaller values of $k_\parallel \rho_i$, the decreasing intrinsic
magnetic helicity $\sigma_m(\V{k})$ from the larger scale modes
reduces the resulting parallel magnetic helicity signature, preventing it
from reaching the observed value $\sigma_m' \simeq-1.0$.
The boundary
associated with $a_1$ is given by the relation $k_{\parallel}
\rho_i =a_1 (k_\perp \rho_i)^{2/3}$ with $a_1=1.162$. The minimum parallel
wavelength is therefore given by $(k_{\parallel} \rho_i)_{{min}} =a_1
(k_\perp \rho_i)_{{min}}^{2/3}= a_1 a_4^{2/3}$.  Tests have shown
that, for a fixed total amount of energy in the parallel modes
relative to the perpendicular modes, the magnetic helicity signature
is relatively insensitive to variations in the perpendicular limits
$a_2$ and $a_4$ as long as $a_2 \le 1.0$.
Unlike the perpendicular limits, the 
minimum and maximum parallel scale lengths are
constrained by the maximum and minimum periods of the observed parallel 
signature from \emph{in situ} magnetic helicity measurements,
shown for instance in Figure~\ref{fig:sept17}. 
Therefore, we choose to fix the
values $a_1=1.162$ and $a_2 = 1.0$, and allow $a_3$ to determine
$(k_\parallel \rho_i)_{{max}}$ and $a_4$ to determine $(k_{\parallel}
\rho_i)_{{min}}$. For
improved clarity in describing our turbulence models, we quote the
more physically relevant values of $(k_{\parallel} \rho_i)_{{min}}$ and
$(k_\parallel \rho_i)_{{max}}$ instead of $a_3$ and $a_4$. These parallel
wavelength limits are informed by the range of $k_\parallel \rho_i$ over
which the EMIC and parallel firehose instabilities arise,
approximately from $k_\parallel \rho_i \in [0.2,2.0]$, as seen in
Figure~\ref{fig:unstable}. The precise $k_\parallel \rho_i$  boundaries 
used for each calculation are stated in the following section.

The energy spectrum of the parallel modes, if this component is
subdominant to the anisotropic \Alfvenic cascade, is not well
constrained by observations.  Since spectra in the direction parallel
to the local mean magnetic field in the solar wind are observed to
have a steeper $-2$ spectral index \citep{Podesta:2009,Wicks:2010,
Chen:2012b,Horbury:2012},
we impose that the magnetic energy spectrum of the parallel modes is a
broken power-law with $k^{-2}$ in the inertial range, transitioning at
$k\rho_i=1$ to $k^{-3.2}$ in the dissipation range. 
The dissipation range scaling for the parallel modes was chosen
to match solar wind observations \citep{Chen:2010b}. Minor variation
of this scaling has no qualitative effect on the resulting synthetic
magnetic helicity measurements. 

An important physical property is that all of the waves populating our
synthetic plasma volumes---including \Alfven and kinetic \Alfven waves
with nearly perpendicular wavevectors, $k_\perp \gg k_\parallel$, and
Alfv\'en, ion cyclotron, fast, and whistler waves with nearly parallel
wavevectors, $k_\parallel \gg k_\perp$---propagate with group
velocities nearly exactly along the local mean magnetic
field. Therefore, two important unknown parameters in our turbulence
model are the parallel Poynting fluxes of electromagnetic wave energy
along the local mean magnetic field for the perpendicular modes
$S_z^\perp$ and for the parallel modes $S_z^{\parallel}$. For the
synthetic spacecraft data method, we define the \emph{normalized
parallel Poynting flux}, $S_z \equiv \hat{\mathbf{z}} \cdot
\V{E}\times\V{B} /|E||B|$, where $\hat{\mathbf{z}}$ is aligned with 
the local mean magnetic 
field $\V{B_0}$. In MHD theory, the cross helicity $\sigma_c$ is
commonly used to specify this balance of upward versus downward wave
energy fluxes along the mean magnetic field, but this standard
definition breaks down at sub-Larmor scales, $k \rho_i \gtrsim
1$, due to the decoupling of the ion and electron fluids and
the introduction of finite magnetic compressibility.
Therefore, we use the more generally applicable normalized
parallel Poynting flux $S_z$ to specify the \emph{imbalance} of upward
versus downward wave energy fluxes. The values of $S_z^\perp$ and
$S_z^\parallel$ used in our turbulence model are constant and
independent of wavevector.

We denote a case with equal
counterpropagating wave energy fluxes as \emph{balanced} turbulence,
and a case with unequal wave energy fluxes as \emph{imbalanced}
turbulence
\citep{Lithwick:2007,Beresnyak:2008,Chandran:2008b,Perez:2009,Podesta:2009}.
Theoretical considerations \citep{Podesta:2011b} suggest that the
parallel modes should be close to unidirectional, $S_z^{\parallel}=\pm
1$, to maximize the parallel magnetic helicity signal. 
We do not assume a balanced or imbalanced distribution of waves
\emph{a priori}, but rather vary the power in sunward and anti-sunward waves
individually for the perpendicular and parallel populations 
to find the best agreement with observations.
We shall see
that the values of $S_z^{\parallel}$ and $S_z^\perp$ in our model are
tightly constrained by comparison to the magnetic helicity
observations.

The final parameter of our model is the relative power between the
perpendicular and parallel populations of waves, parametrized by
$P_\perp/(P_\perp +P_\parallel)=P_\perp/P_{tot}$. Since the parallel
wave modes are expected to be locally concentrated around $k_\parallel
\rho_i \sim 1$, whereas the perpendicular modes are part of an
anisotropic turbulent cascade from large to small scales, we compare
only the power in a band of wavenumbers with 
$k \rho_i \in [0.9,1.1]$.
We shall see that the parameter $P_\perp/P_{tot}$ is also tightly
constrained by these comparisons to the magnetic helicity observations.

In summary, the successful application of the synthetic spacecraft
data method requires that a realistic model of the turbulent
fluctuations can be constructed. Based on theoretical arguments, the
results of direct numerical simulations, and \emph{in situ} solar wind
measurements, we have defined in this section a physically motivated
model for the turbulent fluctuations.  Although the turbulence model
depends on a number of parameters, many of these parameters are well
constrained by observations.  For all of the tests presented in this
study, we choose $k_0 \rho_i=10^{-3}$, $a_1=1.162$, and $a_2=1.0$.
This leaves five free parameters for each turbulence model: (1) the
minimum parallel wavenumber of the parallel waves, $(k_{\parallel}
\rho_i)_{{min}}$; (2) the maximum parallel wavenumber of the parallel
waves, $(k_\parallel \rho_i)_{{max}}$; (3) the imbalance of wave energy
flux for the parallel waves, $S_z^{\parallel}$; (4) the imbalance of
wave energy flux for the perpendicular waves, $S_z^\perp$; and, (5)
the ratio of power in the perpendicular modes to the total power,
$P_\perp/P_{tot}$. As will be demonstrated, even though this is a
large number of parameters, the characteristics of the plot of
magnetic helicity as a function of angle and period strongly constrain
each of these parameters, yielding a valuable new approach to
constrain the properties of the turbulent fluctuations in the solar
wind.



\section{Results of Synthetic Spacecraft Data Method}
\label{sec:Synth}



To generate a synthetic plot of the magnetic helicity as a function of
angle and period to compare directly to spacecraft observations, we
employ the \emph{synthetic spacecraft data method} \citep{Klein:2012}.
The general technique employed to construct synthetic data, outlined
in \cite{Klein:2012}, is briefly summarized here. In this method, first,
the turbulent plasma is modeled by a three-dimensional synthetic
plasma volume containing a uniform background magnetic field $B_0$
and a population of randomly phased linear waves with wavevector distribution
defined by the
turbulence model described in Section~\ref{sec:turbulence}.  The
linear kinetic eigenfunctions of the wave modes are used to specify
all of the fields associated with the turbulence, including the
magnetic field, the electric field, plasma fluid velocity, and the
density.  Next, the plasma is sampled at the position of a probe
moving with uniform velocity $-\V{V}_{SW}$ with respect to the plasma 
to generate reduced time series
comparable to single-point spacecraft measurements. Finally, these
synthetic time series may then be analyzed with procedures identical
to those used to analyze \emph{in situ} spacecraft measurements.

\subsection{Construction of Synthetic Spacecraft Data}

We here focus on the resulting synthetic time series for the magnetic
field using the turbulence model outlined in
Section~\ref{sec:turbulence}.  For each of the three chosen solar wind
intervals, defined in Table~\ref{tab:intervals}, we specify the
synthetic plasma volume by a $256^3$ grid in $(k_x,k_y,k_z)$ space,
where the local mean magnetic field $\V{B}_0$ is aligned with the
$\hat{\mathbf{z}}$ direction. To compare with the \emph{in situ}
measurements, we require a wavevector range that yields spacecraft
frame periods down to $T_{min}=1$~s, so our maximum wavevector
magnitude is computed using $(k\rho_i)_{max}= 2 \pi \rho_i/(T_{min}
V_{SW})$.  For interval \textbf{I}, $\rho_i/V_{SW}=0.544$~s, yielding a
three-dimensional wavevector grid with $0.0267 \le k_j \rho_i \le
3.418$, where $j=x,y,z$.  The other two intervals have similarly
spaced grids.

The eigenfunctions of the chosen wave modes are calculated using the
linear Vlasov-Maxwell dispersion relation \citep{Stix:1992,Quataert:1998}
for each fully resolved wavevector $(k_x,k_y,k_z)$ satisfying $|k_x|,|k_y|
,|k_z| < |k_{max}|$.  
The value of $\beta_i$ is given
by Table~\ref{tab:intervals}, and we take $T_i/T_e=1$ and
$v_{t_i}/c=10^{-4}$. The perpendicular region in
Figure~\ref{fig:region} is populated with randomly phased \Alfven and
kinetic \Alfven waves, and the parallel region is populated with a randomly
phased distribution of either ICWs or whistler waves. The properties
of the turbulence model are detailed in Section~\ref{sec:turbulence}.

The synthetic plasma volume is sampled by a probe with velocity
$-\V{V}_{SW}$ along a trajectory with constant angle $\theta$ 
between $\V{V}_{SW}$ and $\V{B}_0$ to create
a single-point time series corresponding to \textit{in situ} satellite
measurements of the solar wind. 
For this study, we adopt Taylor's
hypothesis and therefore we need not evolve the synthetic plasma
fluctuations in time. 
The synthetic magnetic field
time series for a probe traveling along the trajectory
$\V{r}=\V{V}_{SW}t$ is given explicitly by
\begin{equation}
\V{B}(t_n)  
= \sum_m \sum_{\V{k}} \V{B}_{m}(\V{k})
e^{-i[\V{k} \cdot \V{V}_{SW} t_n - \phi_{m,\V{k}}]},
\label{eqn:TaylorEvolution}
\end{equation}
where the index $m$ represents the distinct contribution
from the fast or \Alfven modes and $\V{k}$ indicates the
contribution from each point on the synthetic plasma grid.
$\phi_{m,\V{k}}$ is the unique random phase assigned to each wave type for
each wavevector on the grid and $\V{B}_{m}(\V{k})$ represents 
the complex Fourier coefficient of wave type $m$ at point $\V{k}$.
The resulting field satisfies a reality condition 
$\V{B}(\V{k})\exp(-i\phi_{m,\V{k}})=\V{B}^*(-\V{k})\exp(+i\phi_{m,\V{-k}})$.
The use of $t_n=n \Delta t$ makes explicit the discrete nature
of the time series. In this work, $\Delta t$ is set to either
$0.5$ s or $1.0$ s.
A detailed discussion of the derivation of
Equation~\ref{eqn:TaylorEvolution}
can be found in Section 3.2.2 of \cite{Klein:2012}.

Once the synthetic data is created, we shift the
time series of magnetic field measurements into RTN coordinates, where
R is anti-parallel to the probe velocity, and T and N define the plane
perpendicular to R, such that $e_R\times e_T=e_N$
where $e_j$ is the unit vector parallel to $R, \ T,$ or $N$.
Next, we calculate the magnetic helicity
$\sigma_m'$:
\begin{equation}
\sigma_m'(f_p)=\frac{2}{|\hat{B}(f_p)|^2}
\mathrm{Im}(\hat{B}_T(f_p) \hat{B}_N^*(f_p)),
\label{eqn:sig_m}
\end{equation}
where the discrete Fourier transform $\hat{B_j}$ is 
defined for frequencies $f_p=p/(n_t \Delta t), p=0,1,...n_t/2$ as
\begin{equation}
\hat{B}_j(f_p)=\sum_{n=0}^{n_t-1}B_j(t_n) e^{2\pi i n p/n_t},
\label{eqn:dft}
\end{equation}
where $j$ corresponds to the $R$, $T$, or $N$ component
of the magnetic field
and $\V{B}(t_n)$ is the synthetic
magnetic time series defined in Equation~\ref{eqn:TaylorEvolution}.
Values for magnetic helicity derived from Equation~\ref{eqn:sig_m}
are independent of our choice for the orientations of $N$ and $T$.

The magnetic helicity $\sigma_m'$ is a function of frequency $f$,
and angle $\theta$. The number of
time steps in each time series is $n_t=128$.  For each synthetic plasma instance
the volume is sampled along 15 trajectories with evenly spaced $\theta$
ranging from $3^\circ$ to $87^\circ$.
This sampling is repeated for an ensemble of 128 independent
runs, over which the values of $\sigma_m'(f,\theta)$ are averaged. 
The size of $n_t$ is selected so that we may sample
over the higher end of the frequency sampled by the observations
The ensemble approach is taken, rather than simply increasing the
size of $n_t$, so that the new phases $\phi_{mj}$ for each realization of
the synthetic plasma volume may stand in for the evolution of 
the turbulent solar wind plasma. The
results are plotted as functions of $T=1/f$ for direct comparison to
the plots of magnetic helicity as a function of angle and period
constructed from \textit{in situ} measurements.
It is shown in the Appendix that the magnetic helicity
maps are all symmetric about $\theta=90^\circ$. Therefore, we only 
need to plot $\sigma_m'$ with $\theta<90^\circ$ in our magnetic helicity maps
generated from the synthetic spacecraft time series,
Figures~\ref{fig:single}-\ref{fig:alias}.

The plasma parameters and the slope of the magnetic power spectrum are
fixed by the observational constraints, so there remain only the five
free parameters of the synthetic magnetic helicity plots:
$(k_{\parallel} \rho_i)_{{min}}$, $(k_\parallel \rho_i)_{{max}}$,
$S_z^{\parallel}$, $S_z^\perp$, and $P_\perp/P_{tot}$.  Although this
is a large number of parameters, the qualitative features of the
observed magnetic helicity plots are affected by variation in each of
these parameters in unique ways, so it is possible to determine quite
strict constraints on all of these parameters. The resulting
constraints provide significant information about the the nature of
the underlying turbulent fluctuations in the solar wind.

\subsection{Single Wave Type Tests}

\begin{figure} [t]
\begin{center}
\includegraphics[width=8.0cm,viewport= 45 0 500 385, clip=true]
{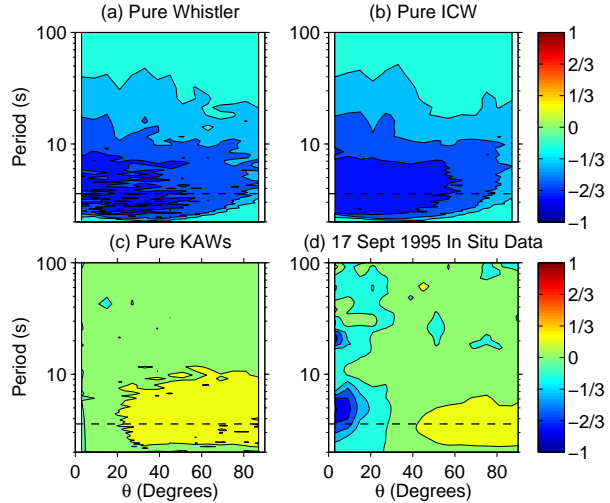}
\caption{$\sigma_m'$ contour maps from synthetic data sets derived 
using only a single mode population.  None of the three solitary mode
types, whistlers with $S_z^\parallel=-1.0$ (panel (a)), ICWs with
$S_z^\parallel=1.0$ (b), or KAWs with $S_z^\perp=0.7$ (c), 
fit the interval \textbf{I} \textit{in situ} magnetic helicity
measurements (d).  The correct sign for the the parallel
and perpendicular populations is reproduced, but the angular extent of
all three signatures are too great.  The dashed black line indicates
the approximate period for $k \rho_i=1$.}
\label{fig:single}
\end{center}
\end{figure}

To gain intuition about the qualitative characteristics of the
magnetic helicity plot due to the different types of waves included in
the turbulence model described in Section~\ref{sec:turbulence}, we
first attempt to fit the solar wind measurements with just one type of
wave: only parallel whistler waves, only parallel ICWs, and only
perpendicular KAWs. The results of these tests for the plasma
parameters of interval \textbf{I} are shown in
Figure~\ref{fig:single}, with the following four panels: (a) only
parallel whistler waves with $S_z^\parallel =-1.0$, (b) only parallel
ICWs with $S_z^\parallel=1.0$, (c) only a critically balanced
distribution of
\Alfven and kinetic \Alfven waves with  $S_z^\perp=0.7$, and (d) solar wind observations from 
 interval \textbf{I}. 

For the two cases (a and b) with only parallel wave modes, the
parallel signatures on the resulting magnetic helicity plots have the
right sign and magnitude, but are much too extended in $\theta$.
The values of $(k_\parallel \rho_i)_{min}=0.428$ and $(k_\parallel
\rho_i)_{max}=1.4$ used for both of these parallel wave cases do lead
to a parallel signature that is limited to the appropriate range in
period $T$.

For case (c) with only perpendicular waves, to achieve the right
magnitude of the perpendicular signature, the wave energy flux cannot
be unidirectional; instead, a value of $S_z^\perp=0.7$ provides the
best fit in magnitude. The angular extent of the perpendicular
signature for this case extends down to $\theta\simeq 20^\circ$,
somewhat more extended than the observed perpendicular signature in
panel (d) that extends only down to $\theta\simeq 40^\circ$.  The
perpendicular signature is naturally limited to periods $T\lesssim
8$~s due to the fact that, for $k_\perp \rho_i \ll 1$, the \Alfven
waves have $\sigma_m(\V{k})\simeq 0$, as shown in
Figure~\ref{fig:sigma_k}. The decrease in positive magnetic helicity
at $T\lesssim 3$~s is an effect of aliasing, as will be demonstrated
in Section~\ref{sec:alias}.

In summary, these tests demonstrate that the data cannot be reproduced
using only a single type of wave.  A combination of both parallel and
perpendicular distributions of waves is necessary to reproduce
qualitatively the parallel and perpendicular signatures and to limit
the extent of these signatures in angle $\theta$ and period $T$.

\subsection{Optimized Models for Intervals \textbf{I}--\textbf{III}}
\label{sec:fit}

\begin{figure*}[t]
\begin{center}
\includegraphics[width=13cm,viewport= 25 0 515 425, clip=true]
{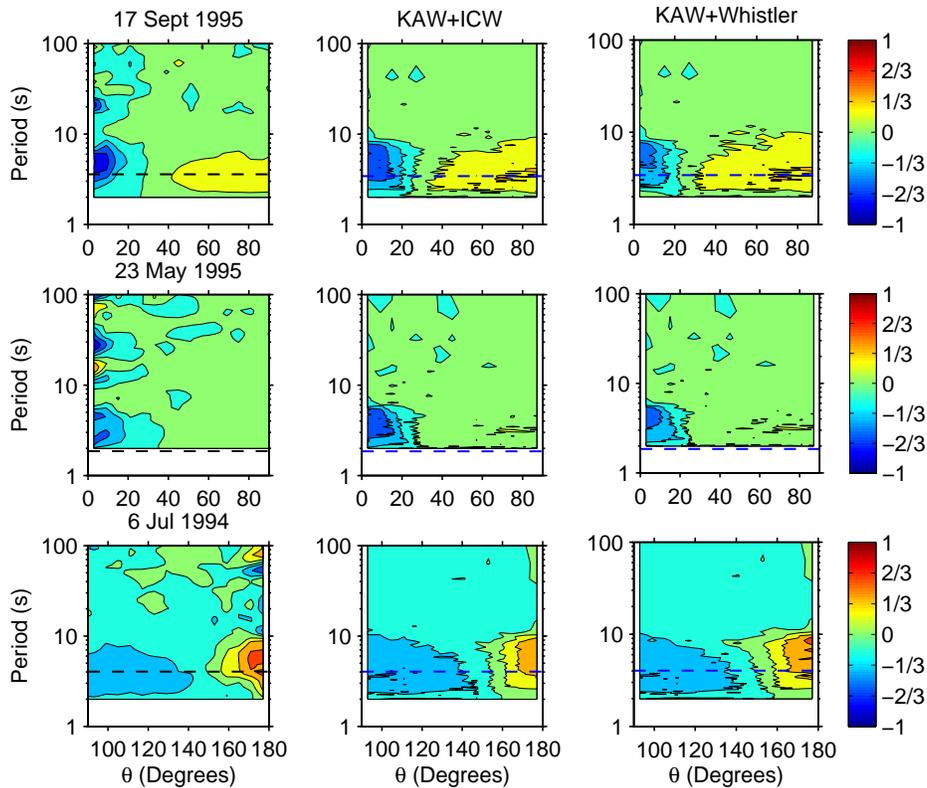}
\caption{Contour plots for $\sigma_m'$ from \textit{in situ} measurements (left column), 
synthetic data generated with ICWs and KAWs (center column) and
synthetic data generated with whistler waves and KAWs (right
column). The rows correspond to the three intervals outlined in the
text: \textbf{I} Sept 17 (top), \textbf{II} May 23 (middle), and
\textbf{III} July 6 (bottom). Parameter values for each of these best fit models are 
given in Tables~\ref{tab:fit_ICW} and~\ref{tab:fit_whistler}.  The dashed
horizontal lines indicate an approximate value for $k \rho_i=1$.}
\label{fig:grid}
\end{center}
\end{figure*}

In this section, we present the optimized turbulence models that lead
to the best fit of the plot of magnetic helicity as a function of angle
and period for the three intervals of Ulysses data described in
Section~\ref{sec:inSitu}. The sensitivity of the result to variations
in each of the model parameters is discussed in detail in
Section~\ref{sec:sens}. 

\begin{table*}[t]
\begin{center}
\begin{tabular}{|c|c||c|c|c|c|c||}
\hline
Interval &  KAW + ICW  & $P_\perp / P_{tot}$ & $S_Z^\perp$ & $S_Z^\parallel$ & $k_\parallel \rho_i ^{min}$ & $k_\parallel \rho_i ^{max}$ \\
\hline
\textbf{I} &Sept 17 & $0.93^{+0.05}_{-0.05}$ & $0.5^{+0.2}_{-0.2}$ &$1.0^{+0.00}_{-0.25}$ &$0.43^{+0.1}_{-0.1}$ &$1.4^{+0.2}_{-0.2}$ \\
\textbf{II} &May 23  & $0.93^{+0.02}_{-0.02}$ & $0.2^{+0.2}_{-0.1}$ &$1.0^{+0.00}_{-0.25}$ &$0.32^{+0.1}_{-0.1}$ &$1.0^{+0.2}_{-0.2}$ \\
\textbf{III} &July 6  & $0.96^{+0.02}_{-0.05}$ & $-0.6^{+0.2}_{-0.2}$ &$-1.0^{+0.25}_{-0.00}$ &$0.42^{+0.1}_{-0.1}$ &$1.6^{+0.2}_{-0.2}$ \\
\hline
\end{tabular}
\caption{Table of parameters for optimized turbulence models
using a combination of ICWs  and KAWs for the three selected intervals.
\label{tab:fit_ICW}}
\end{center}
\end{table*}

\begin{table*}[t]
\begin{center}
\begin{tabular}{|c|c||c|c|c|c|c||}
\hline
Interval &   KAW + Whistler & $P_\perp / P_{tot}$ & $S_Z^\perp$ & $S_Z^\parallel$ & $k_\parallel \rho_i ^{min}$ & $k_\parallel \rho_i ^{max}$ \\
\hline
\textbf{I} &Sept 17 & $0.93^{+0.05}_{-0.05}$ & $0.5^{+0.2}_{-0.2}$ &$-1.0^{+0.25}_{-0.00}$ &$0.43^{+0.1}_{-0.1}$ &$1.4^{+0.2}_{-0.2}$ \\
\textbf{II} &May 23  & $0.92^{+0.02}_{-0.02}$ & $0.2^{+0.2}_{-0.1}$ &$-1.0^{+0.25}_{-0.00}$ &$0.32^{+0.1}_{-0.1}$ &$1.0^{+0.2}_{-0.2}$ \\
\textbf{III} &July 6  & $0.97^{+0.01}_{-0.05}$ & $-0.6^{+0.2}_{-0.2}$ &$1.0^{+0.0}_{-0.25}$ &$0.42^{+0.1}_{-0.1}$ &$1.6^{+0.2}_{-0.2}$ \\
\hline
\end{tabular}
\caption{Table of parameters for optimized turbulence models
using a combination of whistler and KAWs for the three selected intervals.
\label{tab:fit_whistler}}
\end{center}
\end{table*}

A comparison between plots of $\sigma_m'$ as a function of angle
$\theta$ and period $T$ derived from observational data (left
column) and from synthetic spacecraft data using either parallel ICWs
(center column) or parallel whistler waves (right column) is shown in
Figure~\ref{fig:grid} for the three intervals.  
The values yielding
the best fit for each of the parameters are presented in
Table~\ref{tab:fit_ICW} for models using parallel ICWs and in
Table~\ref{tab:fit_whistler} for models using parallel whistler
waves. An important inclusion in these tables is the range of
variation of each of these values within which the fit did not degrade
noticeably.

For each interval, the five adjustable parameters, $P_\perp/P_{tot}$,
$S_z^\perp$, $S_z^{\parallel}$, $(k_{\parallel} \rho_i)_{{min}}$, and
$(k_\parallel \rho_i)_{{max}}$, were varied to obtain the best
correspondence by eye between the observed and synthetic magnetic
helicity plots. The value of $S_z^\parallel$ was chosen to be
unidirectional, $S_z^\parallel= \pm 1.0$ to yield the maximum
amplitude signal for the parallel signature. The values of
$(k_{\parallel} \rho_i)_{{min}}$ and $(k_\parallel \rho_i)_{{max}}$ where
chosen to match the maximum and minimum periods of the dominant
parallel signature. The value of $P_\perp/P_T$ was varied to match the
angular extent of the perpendicular signature with the spacecraft
measurements while still allowing a parallel structure with
$|\sigma_m'| \ge 0.8$. The value of $S_z^\perp$ was adjusted so that
the amplitude of the perpendicular signature would match that of the
\textit{in situ} measurement.  The correspondence between both our
models, using either ICWs or whistlers as the parallel component, and
the \textit{in situ} measurements is quite good for all three
intervals.

The stringent constraints on the parameters of the turbulence model,
particularly the power ratio $P_\perp/P_{tot}$, provides important
information about the nature of turbulence in the solar wind.

\subsection{Sensitivity to Parameters}
\label{sec:sens}
\begin{figure*} [t]
\begin{center}
\includegraphics[width=13cm,viewport= 25 0 490 425, clip=true]
{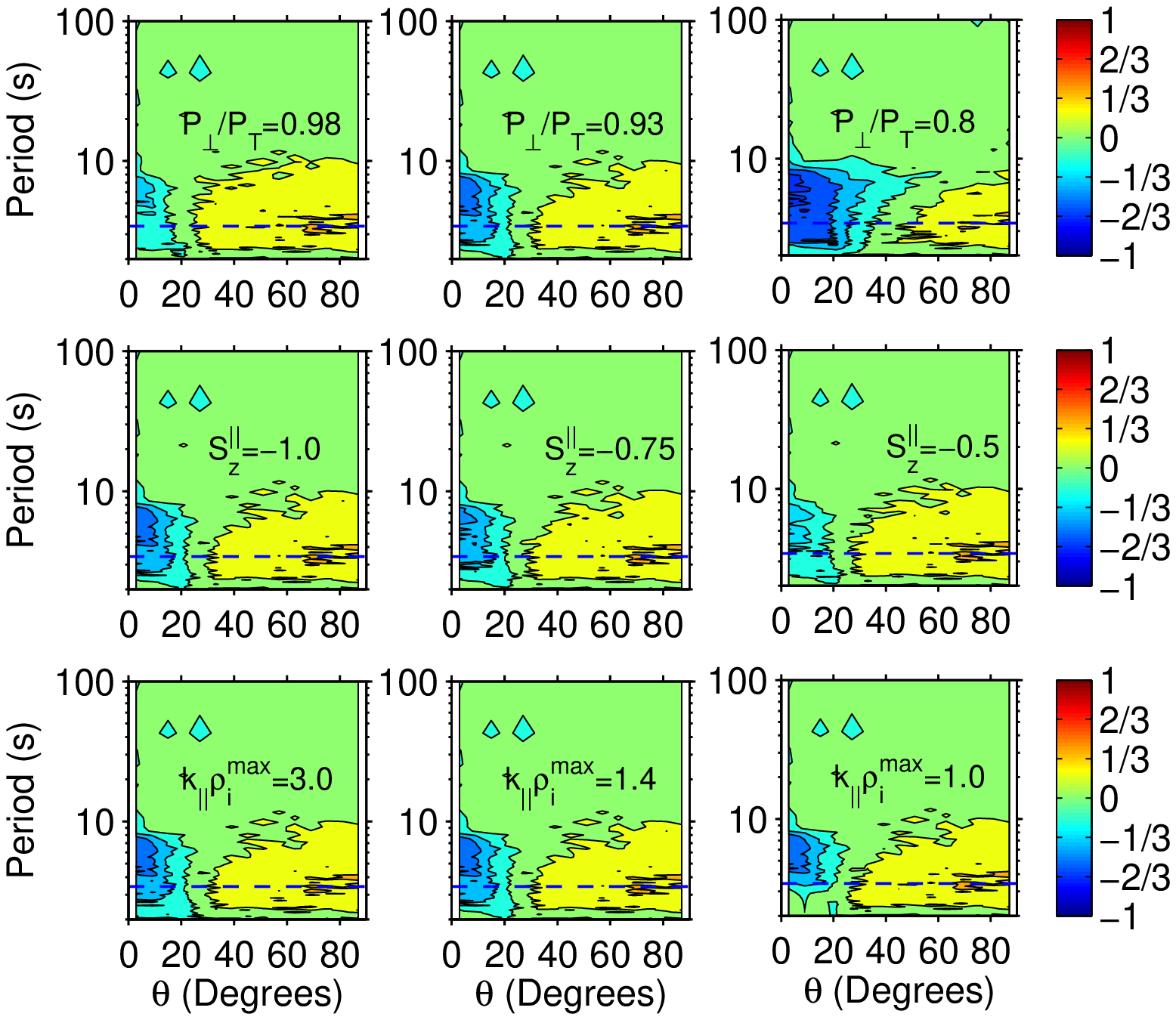}
\caption{Contours of $\sigma_m'$  for a variation of $P_\perp/P_{tot}$ (top row), 
$S_z^{\parallel}$ (center row) and $k_\parallel \rho_i^{max}$ (bottom row).
The plasma parameters are set to mimic interval \textbf{I} conditions.
All nine cases are mixtures of whistler waves and KAW; similar results
(not shown) are obtained for ICW and KAW combinations. The dashed horizontal lines
indicate approximate values for $k \rho_i=1$.}
\label{fig:sens}
\end{center}
\end{figure*}

It is worthwhile demonstrating the sensitivity of the 
synthetic magnetic
helicity plot to variations in the five adjustable parameters:
$P_\perp/P_{tot}$, $S_z^\perp$, $S_z^{\parallel}$, $(k_{\parallel}
\rho_i)_{{min}}$, and $(k_\parallel \rho_i)_{{max}}$. Variation in each of
these parameters produces a unique effect on the $\sigma_m'$ plot,
allowing these five parameters to be rather strictly constrained
without significant problems of degeneracy.  

The variations of the five parameters in this section are performed about
the best fit values for interval \textbf{I} for combinations of KAW and 
whistler waves. Qualitatively similar results are found when combining
KAW and ICWs. The fiducial values for the parameters, shown
in Table~\ref{tab:fit_whistler}, are $P_\perp/P_{tot}=0.93$,
$S_Z^\perp=0.5$, $S_Z^\parallel=-1.0$, $(k_\parallel \rho_i )^{min}=0.43$,
and $(k_\parallel \rho_i )^{max}=1.4$.

The angular extent of the parallel and perpendicular signatures on the
observed magnetic helicity plot strongly constrains the relative power
between the parallel and perpendicular waves, $P_\perp/P_{tot}$.  In
the top row of Figure~\ref{fig:sens}, we plot the result of variations
in $P_\perp/P_{tot}$ for interval \textbf{I} over the range $0.8 \le
P_\perp/P_{tot} \le 0.98$ for a combination of parallel whistler waves
and perpendicular KAWs. Similar results are obtained if one chooses
ICWs instead of whistler waves for the parallel component of the
turbulence model. The results in Figure~\ref{fig:sens} demonstrate
that, for $P_\perp/P_{tot}\lesssim 0.9$, the parallel signature of
$\sigma_m'$ is too extended and the perpendicular signature is too
limited in angular range to match the observations. For the three
intervals studied here, $P_\perp/P_{tot}$ varies from $0.97$ to $0.92$
(see Tables~\ref{tab:fit_ICW} and~\ref{tab:fit_whistler} for the best
fit values and their acceptable ranges), but in all cases a majority
of the total power is found in the perpendicular wave modes, with only
a small fraction of the total wave energy attributed to the parallel
wave modes.
 
The sign and amplitude of $\sigma_m'$ for the perpendicular signature
determines the allowable range of values for $S_z^\perp$.  The sign of
$S_z^\perp$ is simply given by the sign of $\sigma_m'$.  The amplitude
of $\sigma_m'$ for the perpendicular signature allows one to determine
the balance of the wave energy fluxes along the local mean magnetic
field.  Specifying a population of unidirectional KAWs with
$|S_z^\perp|\simeq 1$ yields a magnitude of $\sigma_m'$ that is too
large, while for a balanced case with $|S_z^\perp|\simeq 0$, the
perpendicular signature disappears. These cases are clearly
demonstrated by comparing the best fit plots in Figure~\ref{fig:grid}.
The top and bottom rows are rather imbalanced, with $S_z^\perp=0.7$
and $S_z^\perp=-0.6$, respectively. The middle row demonstrates a very
important feature of the magnetic helicity plot. In this case, the KAW
component of the turbulence is much more balanced, with
$S_z^\perp=0.2$.  Although the perpendicular signature has vanished,
the angular extent of parallel signature is still restricted to small
$\theta$. In the case where there is no perpendicular wave energy
at all, as shown in the pure parallel cases highlighted in
Figure~\ref{fig:single}, the parallel signature extends to large
angles.  Therefore, the absence of a perpendicular signature in the
magnetic helicity plot does not indicate the absence of a
perpendicular KAW turbulent cascade, but rather a case of nearly balanced
KAW turbulence.

A brief discussion in \cite{Podesta:2011a} commented on the possibility
that the symmetry of the magnetic helicity signature
about $\theta=90^\circ$ observed in the solar wind could be indicative
of a balanced population of KAWs. 
As a balanced population of
KAWs would produce a magnetic helicity of zero, they suggested
other mechanisms may be needed to account for the observed nonzero
magnetic helicity signature near $\theta=90^\circ$. However,
the concerns raised by \cite{Podesta:2011a} are invalid.
In the Appendix, we prove that the magnetic helicity
map is symmetric about $\theta=90^\circ$. This proof differs from the
one presented by \cite{He:2012}. Our work with the synthetic time series
shows that an imbalanced population
of KAWs generates a normalized helicity signature that is symmetric about
$\theta=90^\circ$ and is consistent with solar wind observations. This minor 
comment by \cite{Podesta:2011a} is erroneous and we feel obliged to correct that 
mistake in this paper.

Analogous to the manner in which the perpendicular magnetic helicity
signature diminishes as the wave energy fluxes of KAWs become more
balanced, $S_z^\perp \rightarrow 0$, the amplitude of the parallel
$\sigma_m'$ signature diminishes as $|S_z^\parallel|$ is reduced from
$1$ to $0$.  This reduction is shown in the center row of figure
\ref{fig:sens}, where $S_z^\parallel$ for a population of whistler
waves increases from $-1$ through $-0.75$ to $-0.5$. As stated in
previous studies \citep{Podesta:2011b}, the parallel component must
be nearly unidirectional to match the observed amplitude of the
parallel signature in $\sigma_m'$.

Finally, the region of wavevector space filled by the parallel
population of waves is directly constrained by the observed minimum
and maximum period of the parallel signature on the magnetic helicity
plot. The maximum period $T_{max}$ is directly controlled by our model
parameter $(k_{\parallel} \rho_i)_{{min}}$, and the minimum period
$T_{min}$ is directly controlled by $(k_\parallel \rho_i)_{{max}}$.
The observed maximum period of the parallel signature\footnote{Note
that we focus here on the largest peak of the parallel signature at
$T< 10$~s. The signal of this largest peak is clearly significant, but
it is not clear if the smaller peaks that sometimes appear at $T>
10$~s are significant.} for intervals \textbf{I}--\textbf{III}, as seen
in the first column in Figure~\ref{fig:grid}, has a range of values
6~s $<T_{max}<10$~s. These values are matched in the turbulence models
in Figure~\ref{fig:grid}---both ICW (center column) and whistler
(right column) models---by setting values of the the minimum parallel
wavenumber $(k_{\parallel}\rho_i)_{{min}} =0.43,\ 0.32, \ 0.42$.
Likewise, to see the effect of $(k_\parallel \rho_i)_{{max}}$ on the
observed $T_{min}$ of the parallel magnetic helicity signature, in the
bottom row of Figure~\ref{fig:sens}, we plot $(k_\parallel
\rho_i)_{{max}}=3.0,\ 1.4,\ 1.$ It is clear the increasing
$(k_\parallel \rho_i)_{{max}}$ has the effect of decreasing $T_{min}$
of the parallel signature.  Further discussion of the effect of
aliasing on the $T_{min}$ limit of the parallel signature for the
$(k_\parallel \rho_i)_{{max}}=3.0$ case is presented in
Section~\ref{sec:alias}.

\subsection{Effects of Aliasing}
\label{sec:alias}

\begin{figure} [t]
\begin{center}
\includegraphics[width=8.5cm,viewport= 40 0 500 400,clip=true]
{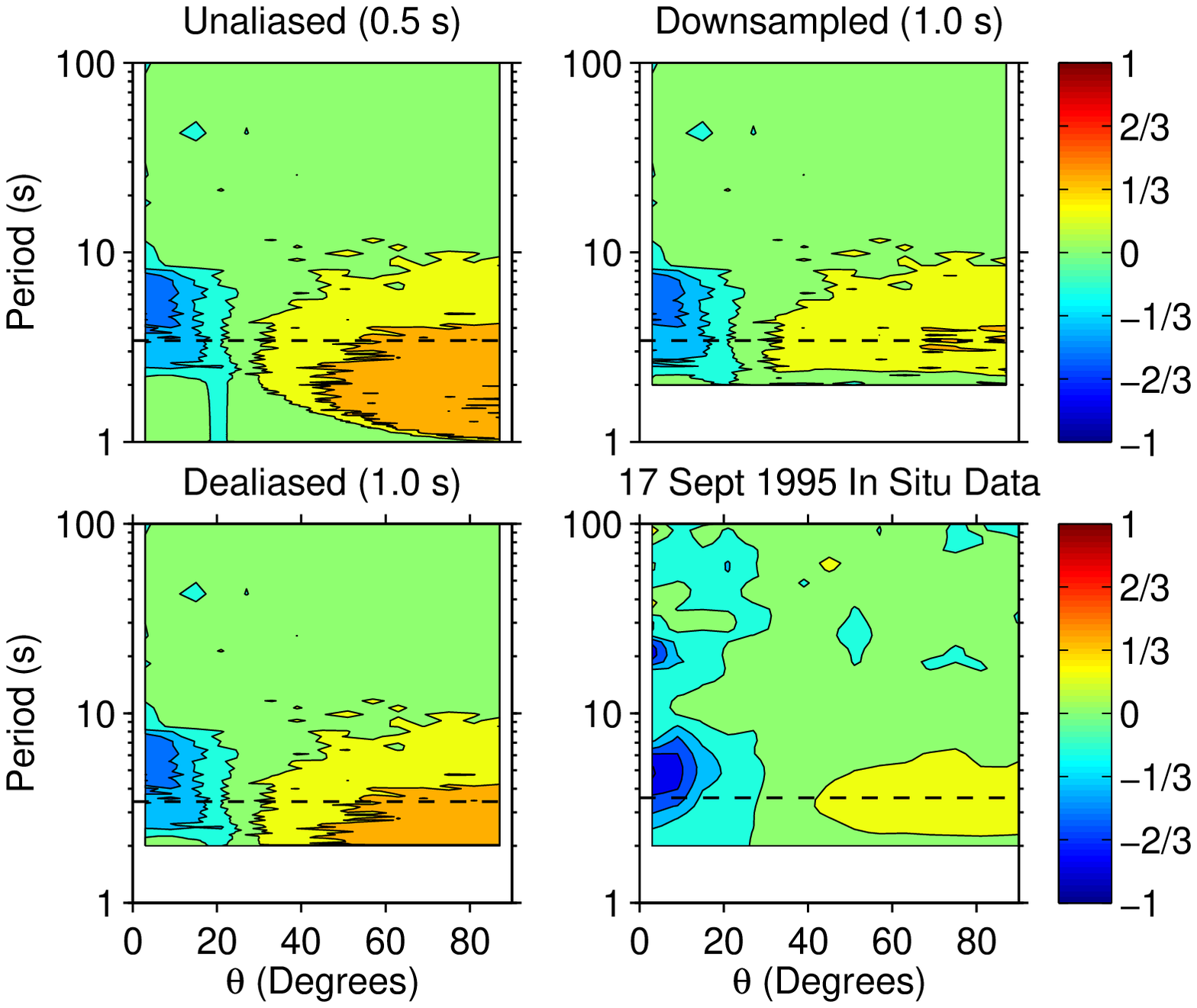}
\caption{The effects of aliasing on the magnetic helicity $\sigma_m'$:
 unaliased signal from the $\Delta t=0.5$~s time series (upper left),
 aliased signal from the downsampled $\Delta t=1$~s time series (upper
 right), the dealiased signal from low-pass filtered, downsampled
 $\Delta t=1$~s time series (lower left), and \emph{Ulysses}
 observations from interval \textbf{I} (lower right).}
\label{fig:alias}
\end{center}
\end{figure}

\begin{figure} [t]
\begin{center}
\resizebox{3.1in}{!}{\includegraphics*[0.3in,6.in][8.0in,9.6in]
{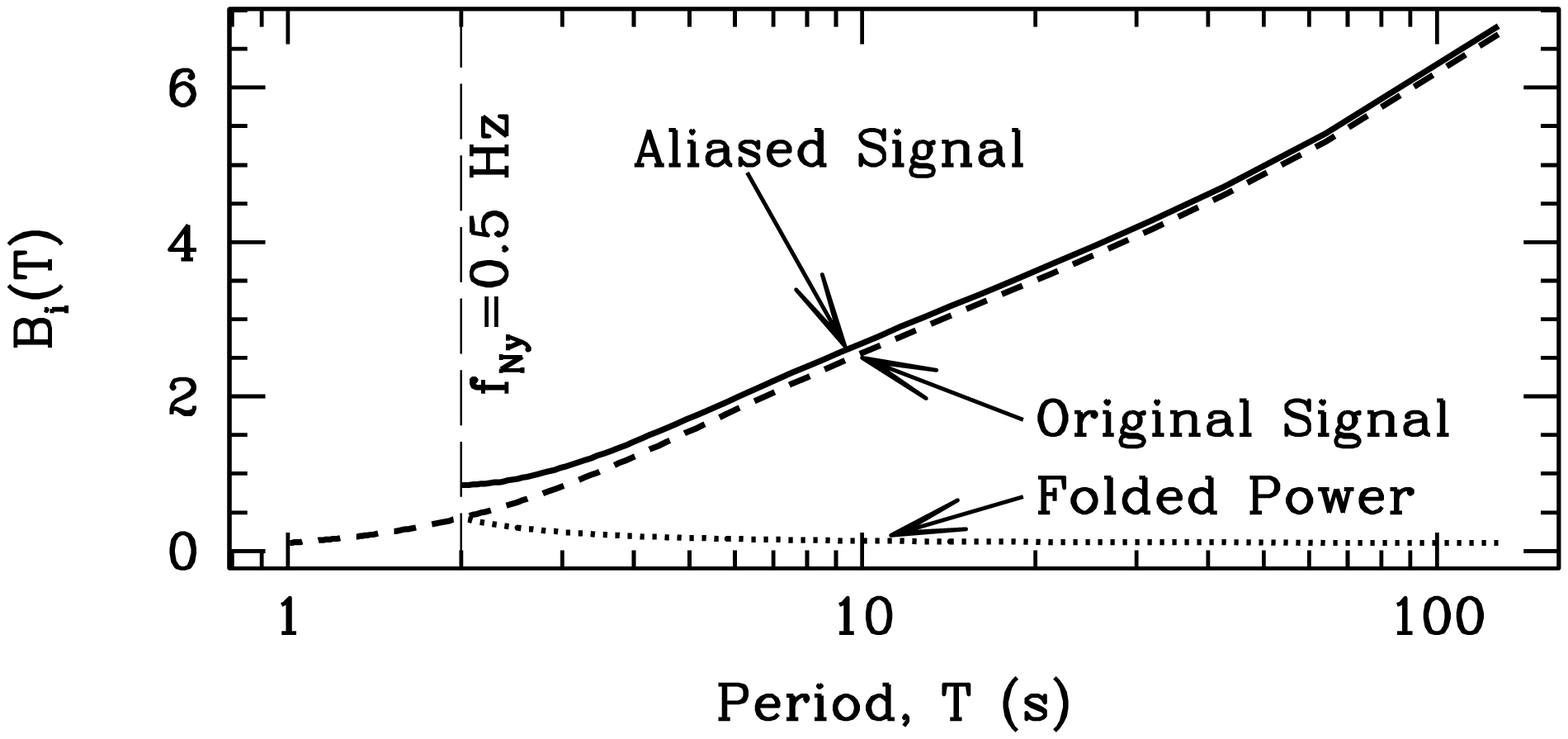}}
\caption{Diagram of how aliasing affects the Fourier transform of a 
magnetic field component, $B_i(T)$. The Nyquist frequency
$f_{Ny}=0.5$~Hz (long dashed) corresponds to a Nyquist period of
$T_{Ny}=2$~s. Power in the original signal (short dashed) at
$T<T_{Ny}$ is folded back (dotted) into the measured range at
$T>T_{Ny}$, leading to an aliased signal (solid) for the Fourier
transform of the magnetic field component, $B_i(T)$.
\label{fig:alias2}}
\end{center}
\end{figure}

A salient feature of the observed plots of magnetic helicity as a
function of angle, $\theta$, and period in the spacecraft frame, $T$,
is that the perpendicular signature drops to zero magnetic helicity,
$\sigma_m' \simeq 0$, at the shortest periods, or highest
frequencies. This has previously been interpreted as indicating that
the anti-sunward and sunward energy fluxes of the waves responsible
for the perpendicular signature are gradually becoming more balanced
at smaller scales \citep{He:2012}.  In stark contrast to this
interpretation, we hypothesize here that the observed drop of the
perpendicular signature to $\sigma_m' \simeq 0$ at the highest
frequencies is simply an artifact of aliasing, and we here present evidence
in support of this hypothesis.

The sampling interval for the \emph{Ulysses} measurements presented in
Figure~\ref{fig:grid} is $\Delta t=1$~s, leading to a Nyquist frequency of
$f_{Ny}=0.5$~Hz. The turbulent magnetic field fluctuations measured in
the solar wind certainly contain power at frequencies $f>f_{Ny}$, so
it is inevitable that aliasing will affect the Fourier transforms of
the measured magnetic field, and this may alter the determination of
the magnetic helicity of the turbulent fluctuations. We explore here
how such aliasing impacts the appearance of the plot of magnetic
helicity as a function of angle and period.

Let us consider how aliasing impacts the magnetic helicity
determination for interval \textbf{I}. Assuming the Taylor Hypothesis
\citep{Taylor:1938}, the maximum frequency $f_{max}$ of the magnetic field  
in our synthetic plasma volume occurs when the probe samples directly
along one of the coordinate directions, $f_{max} = (k \rho_i)_{max}/(2
\pi) (V_{SW}/\rho_i)$. For $(k \rho_i)_{max}=3.418$ and
$\rho_i/V_{SW}=0.544$~s, we obtain $f_{max} =1$~Hz. Note that since we
do not sample exactly along the coordinate directions ($\theta=0$ or
$\theta=\pi/2$), the maximum frequency that we obtain is $f_{max}
<1$~Hz, avoiding the marginally aliased case when sampling at a time
resolution of $\Delta t=0.5$~s.  Therefore, generating synthetic time series
with a time resolution of $\Delta t=0.5$~s is sufficient to capture completely
all information in the signal without aliasing, since the Nyquist
frequency for this sampling rate is $f_{Ny}=1$~Hz and  therefore
$f_{max} < f_{Ny}$. Constructing a magnetic helicity plot using these
$\Delta t=0.5$~s synthetic measurements leads to the unaliased result plotted
in the upper left panel of Figure~\ref{fig:alias}.

If the synthetic data is instead downsampled to a time resolution of
$\Delta t=1$~s (equivalent to simply eliminating every other data
point from the $\Delta t=0.5$~s time series), the resulting Nyquist
frequency is $f_{Ny}=0.5$~Hz, and therefore all power at frequencies
$f>f_{Ny}$ will be folded back into the frequency range $f<f_{Ny}$.
This aliasing is illustrated schematically in Figure~\ref{fig:alias2},
where the Fourier transform of an arbitrary component of the magnetic
field, $B_i$, is plotted not as a function of frequency $f$, but as a
function of period $T=1/f$, to make more direct connection to the
magnetic helicity plot.  The Nyquist frequency $f_{Ny}=0.5$~Hz (long
dashed) therefore corresponds to a Nyquist period of $T_{Ny}=2$~s.  In
this figure, the original turbulent magnetic field signal (short
dashed) contains power below the Nyquist period, $T<T_{Ny}$.  This
power is folded back (dotted), with a 180$^\circ$ phase shift, into
the range $T>T_{Ny}$.  This folded power combines with the original
signal, leading to the aliased Fourier transform, $B_i(T)$ (solid).

Note that the scaling of the original signal as a function of period
in the diagram in Figure~\ref{fig:alias2} corresponds to our
turbulence model as described in Section~\ref{sec:parameters}. This
model yields a magnetic energy frequency spectrum that scales as
$f^{-5/3}$ at $k\rho_i \ll 1$ and as $f^{-2.8}$ at $k\rho_i \gg 1$.
It is important that a realistic frequency scaling is used to
construct the schematic diagram in Figure~\ref{fig:alias2}, because
the scaling of the amplitude of the signal as a function of frequency,
or alternatively as a function of period, determines quantitatively
how much of an impact aliasing will have on the final result.

In Figure~\ref{fig:alias2}, the difference between the aliased signal
(solid) and the original signal (dashed) as a function of period $T$
provides useful guidance to develop intuition about the range of $T$
in which aliasing significantly alters the resulting signal.  In this
case, the difference between the original and aliased signal at the
Nyquist period $T_{Ny}=2$~s is the same magnitude as the original
signal, indicating that aliasing is likely to have a large effect at
this shortest measured period.  The difference remains significant for
periods $T
\lesssim 4$~s, suggesting that aliasing will likely affect the
resulting magnetic helicity plot for all periods shorter than $T\sim
4$~s, supporting our hypothesis.

We confirm the hypothesis that aliasing leads to the reduction of the
magnetic helicity signal at the lowest measured periods using the
synthetic data method, as shown in Figure~\ref{fig:alias}. In the
upper right-hand panel, the magnetic helicity is plotted for the
downsampled $\Delta t=1$~s data. Here we expect the magnetic helicity
to be affected significantly by aliasing for all periods $T \lesssim
2 T_{Ny} =4$~s, as indicated schematically in Figure~\ref{fig:alias2}. One can
clearly see in the upper right-hand panel of Figure~\ref{fig:alias}
that the magnetic helicity of the perpendicular signature indeed drops
to $\sigma_m' \simeq 0$ at the lowest periods, in qualitative
agreement with the observed data for interval \textbf{I} plotted in
the lower right-hand panel, but starkly different from the unaliased
data in the upper left-hand panel.

The effect of aliasing can be eliminated if the original time series
is first passed through a low-pass filter to remove all power at
frequencies $f \ge f_{Ny}$ before the measurements are Fourier
transformed to convert from a time series to a frequency spectrum. In
this case, a low-pass filter was applied to zero out any frequencies
in the range $f \ge 0.5$~Hz.  This filtering was accomplished by the
following procedure: (i) the $\Delta t=0.5$~s time series is Fourier
transformed to frequency; (ii) all Fourier coefficients for $f \ge 0.5
$ Hz are set to zero; (iii) the resulting truncated Fourier frequency
spectrum is transformed back to a time series; and (iv) the resulting
$\Delta t=0.5$~s time series is downsampled to a $\Delta t=1$~s time
series by eliminating every other data point. The magnetic helicity
plot is then constructed from the resulting dealiased time series,
producing the result plotted in the lower left-hand panel of
Figure~\ref{fig:alias}.

A comparison of the four panels plotted in Figure~\ref{fig:alias}
strongly supports our hypothesis that aliasing is responsible for the
drop in the magnetic helicity of the perpendicular signal at the
lowest periods.  The unaliased magnetic helicity plot is presented in
the upper left, created by choosing a sampling time that is short
enough to avoid aliasing, $\Delta t=0.5$~s in this case.  The observed
magnetic helicity plot (lower right) closely resembles the aliased
magnetic helicity plot (upper right) that was constructed from the
downsampled $\Delta t=1$~s time series. The dealiased magnetic
helicity plot (lower left), constructed by low-pass filtering the data
before downsampling, looks significantly different from the
observations (lower right), but is identical to the unaliased result
(upper left) at all periods. 

This comparison strongly supports our hypothesis that aliasing
significantly affects the appearance of the magnetic helicity plot at
the lowest measured periods. The lesson here is that caution must be
exercised  not to over-interpret the apparent return of $\sigma_m'$
to zero at the lowest periods.

Magnetic helicity plots generated from the \emph{STEREO} mission,
presented in Figure~4 of \cite{Podesta:2011a}, further support this
result.  The sampling time of \emph{STEREO A} magnetic field
measurements is $\Delta t=0.125$~s, yielding a Nyquist frequency
$f_{Ny}=4$~Hz.  Although the plasma parameters for this measured
interval are not significantly different from the \emph{Ulysses}
measurements presented here, the non-zero $\sigma_m'$ of the
perpendicular signature extends to much lower periods, $T \gtrsim 0.5
s$. In addition, the perpendicular signature of the \emph{STEREO}
measurements again drops to $\sigma_m' \simeq 0$ at $T \sim 2 T_{Ny} =
0.5$~s, as our intuition about the effect of aliasing from
Figure~\ref{fig:alias2} suggests.

It is worthwhile noting that, for the \emph{STEREO} measurements
above, the low-period cutoff of the parallel signature is a $T\sim
1$~s, a significantly longer period than the Nyquist period,
$T_{Ny}=0.25$~s.  Therefore, these measurements suggest that the
parallel signature cutoff is not due to aliasing, in contrast to the case
for the perpendicular signature.


\section{Discussion}
\label{sec:discussion}


The magnetic helicity of the turbulent fluctuations, plotted as a
function of period and angle, provides a new avenue for exploring the
nature of turbulence in the solar wind.  We have applied the synthetic
spacecraft data method to model \emph{in situ} magnetic helicity observations
and thereby obtained new insights into the nature of the 
underlying plasma dynamics.  Using a
turbulence model inspired by the current theory for anisotropic
\Alfvenic turbulence and by linear kinetic plasma physics, we directly
compare synthetic magnetic helicity plots to those obtained from three
time intervals measured \emph{in situ} by the \emph{Ulysses} spacecraft.
Here we explain how the findings in this study provide support for, or
conflict with, the results of previous related studies.

This work is motivated by two groundbreaking observational studies of
the magnetic helicity of fluctuations in the solar
wind. \citet{He:2011a} were the first to use a wavelet analysis
technique to construct the magnetic helicity of the turbulent fluctuations
as a function of both the period in the spacecraft frame and the angle
between the solar wind flow velocity and the local mean magnetic
field.  This investigation used magnetic field measurements in the
ecliptic from the \emph{STEREO} spacecraft.  For an outward magnetic
sector, this study found a dominant negative magnetic helicity at small
angles $\theta < 30^\circ$, and a broad region of dominantly positive
magnetic helicity at $40^\circ <\theta < 140^\circ$. In this work, we
label the narrow feature with negative magnetic helicity at $\theta <
30^\circ$ as the \emph{parallel signature} and the broad feature with
positive magnetic helicity at $40^\circ < \theta < 140^\circ$ as the
\emph{perpendicular signature}.  Their interpretation was that the
parallel signature arises from ICWs and that the perpendicular
signature is caused by KAWs or whistler waves. They speculated that a
parallel turbulent cascade of Alfv\'en-cyclotron waves may be
responsible for the parallel signature, and that the drop to zero
magnetic helicity in the dissipation range was due to the effect of
pinning, in which the energy fluxes of sunward and anti-sunward waves
gradually become more balanced at smaller scales.

A complementary study by \citet{Podesta:2011a} employed \emph{Ulysses}
measurements at high heliographic latitudes to obtain magnetic
helicity results in accord with the earlier ecliptic observations
\citep{He:2011a}. The interpretation of the results by \citet{Podesta:2011a},
however, differed substantially from that of
\citet{He:2011a}. \citet{Podesta:2011a} suggested that the perpendicular 
signature was most naturally associated with the anisotropic \Alfvenic
turbulent cascade
\citep{Goldreich:1995,Leamon:1999,Bale:2005,Howes:2008a,Howes:2008b,
Schekochihin:2009,Sahraoui:2010,Howes:2011a,Howes:2011b}, consisting of 
\Alfven waves at MHD scales $k \rho_i \ll 1$ and KAWs at kinetic scales
$k_\perp \rho_i \gtrsim 1$. In stark contrast to the interpretation of the 
\citet{He:2011a},  \citet{Podesta:2011a} proposed that the parallel 
signature is not the result of a parallel turbulent cascade, but
rather that it arises from waves driven by kinetic temperature
anisotropy instabilities in the solar wind. 

In a subsequent detailed study, \citet{Podesta:2011b} argued that, due
to the large magnitude of the normalized magnetic helicity associated
with the parallel signature, $|\sigma_m'| \simeq 1$, the waves
responsible must propagate nearly unidirectionally. For outward
magnetic sectors, the negative magnetic helicity of the parallel
signature could be alternatively explained either by (i) ICWs
propagating away from the sun along the interplanetary magnetic field
or by (ii) whistler waves propagating toward the sun along the
interplanetary magnetic field.  They went on to show, in the presence
of a significant differential alpha-proton flow along the magnetic
field $V_d \sim v_A$ (where the field-aligned drift velocity $V_d
\equiv v_\alpha-v_p $), two quantitative results that could explain
the observations: (i) for a temperature anisotropy of the sense $T_{p
\perp}/T_{p \parallel}>1$, the electromagnetic ion cyclotron (EMIC) instability
preferentially generates anti-sunward propagating ICWs; and (ii) for a
temperature anisotropy of the sense $T_{p \perp}/T_{p \parallel}<1$,
the parallel firehose instability preferentially generates sunward
propagating whistler waves.

Finally, \citet{He:2012} employed a simplified analytical model of the
turbulent fluctuations, similar to the approach introduced by
\citet{Howes:2010a}, to explore whether a two-component model of
parallel (slab) and perpendicular (critically balanced) \Alfvenic
fluctuations could explain the qualitative and quantitative features
of the observed magnetic helicity plot. This model included a
distribution of power in wavevector space with both a distribution of
slab (parallel) ICWs and a critically balanced distribution of nearly
perpendicular \Alfven waves and KAWs. In addition, the model included
a prescribed ``imbalance'' function that defined the gradual balancing
of anti-sunward and sunward \Alfven wave energy fluxes, inspired by
the concept of pinning of Elsasser power spectra at the dissipation
scales. Their study concluded that this two-component, slab plus
critical balance model of the turbulence could qualitatively reproduce
the parallel and perpendicular signatures of the observed magnetic
helicity plots. In addition, with appropriate tuning of the imbalance
function, the reduction of the magnetic helicity to $\sigma_m' \simeq
0$ at short periods was also reproduced by the model.

Now we turn our attention to how the method and results of the present
study compare to those of these previous investigations, focusing
initially on the results and interpretation of the perpendicular
signature. One of the questions posed by \citet{Podesta:2011a} was why
the perpendicular signature was so broad, extending over roughly the
angular range $40^\circ < \theta < 140^\circ$, even though
multi-spacecraft measurements of the turbulent fluctuations at these
scales \citep{Sahraoui:2010,Narita:2011,Roberts:2013} show
fluctuations with nearly perpendicular wavevectors, $\theta=90^\circ
\pm 10^\circ$?  The answer found through application of the synthetic
spacecraft data method is that \emph{the angular width of the
perpendicular (and the parallel) signatures is controlled by the
relative power between perpendicular and parallel wave distributions
at the scales associated with spacecraft-frame periods} $1 \mbox{
s}\lesssim T \lesssim 10$~s.  Surprisingly, for a critically balanced
distribution of \Alfven and kinetic \Alfven waves, if they are the
only waves present, the perpendicular signature would extend to even
smaller angles than observed, as shown in panel (c) of
Figure~\ref{fig:single}. Similarly, if only a parallel distribution of
whistler waves or ICWs is present, their parallel signature can extend
all the way out to angles near $\theta \sim 90^\circ$, as shown by
panels (a) and (b) of the same figure. The limitation of the angular
range of both the perpendicular and the parallel signatures is controlled by
the relative power between the perpendicular and parallel waves, as
demonstrated by the top row of Figure~\ref{fig:sens}.

The commonly observed transition at $\theta \sim 30^\circ$ requires
that the perpendicular waves contain more than 90\% of the power in
the wavenumber band $0.9 \le k \rho_i \le 1.1$. For the three
\emph{Ulysses} intervals that we examined, we obtained values of
$P_\perp/P_{tot}$ from 92\% to 97\%, as presented in
Tables~\ref{tab:fit_ICW} and~\ref{tab:fit_whistler}.  This contrasts
significantly with the model used by \citet{He:2012}, in which the parallel
component had 30\% of the total power. It is
worthwhile noting that, due to the statistical averaging of spacecraft
measurements over long time intervals (from one to a few days), the
waves responsible for the parallel signature may indeed have a much
higher relative power, but be spatially intermittent, compared to the
anisotropic \Alfvenic turbulent cascade leading to the perpendicular
signature, which is likely to be space-filling. But, in terms of the
total power of magnetic field fluctuations in the solar wind, our
results suggest that the parallel waves constitute on average about
5\% of the power at scales near $k \rho_i=1.0$.

The magnitude of the magnetic helicity of the perpendicular signature
directly provides information about the balance of anti-sunward and
sunward \Alfven wave energy fluxes, parametrized by the normalized
parallel Poynting flux, $S_z^\perp$. The synthetic spacecraft data
method enables a determination of the imbalance of wave energy fluxes
by adjusting parallel Poynting flux until the magnitude of the
perpendicular signature from the synthetic plots matches that of the
observed plots. This imbalance may vary significantly between
different solar wind intervals, as demonstrated by the three
\emph{Ulysses} investigated here, which yield values for  $S_z^\perp$
of 0.5, 0.2, and $-0.6$, as presented in Tables~\ref{tab:fit_ICW}
and~\ref{tab:fit_whistler}.

Another difference between our study and previous works is the
explanation for the observed drop of the magnetic helicity of the
perpendicular signature to $\sigma_m' \simeq 0$ at the shortest
measured periods. \citet{He:2011a} and \citet{He:2012} suggest that
this is due to a gradual balancing of the anti-sunward and sunward
wave energy fluxes, such that $|S_z^\perp|$ decreases in magnitude
with decreasing length scale. We disagree with this interpretation,
and show evidence that this decrease of $\sigma_m'$ is merely an
artifact of aliasing, as shown in Figures~\ref{fig:alias}
and~\ref{fig:alias2}. 
Therefore, the observations do not necessarily imply that pinning occurs
at kinetic scales. The possibility
that the kinetic turbulence becomes balanced at some point in the
energy cascade has not been incorporated into our model which assumes that 
the imbalance in the energy spectrum is scale independent.

Turning our attention to the parallel signature of the magnetic
helicity plot, we present evidence in
support of the idea that the waves responsible for this feature are
driven by temperature anisotropy instabilities
\citep{Podesta:2011a,Podesta:2011b},
as opposed to the suggestion that a parallel turbulent
cascade is responsible for these modes \citep{He:2011a,He:2012}.  
In particular, when a
field-aligned differential alpha-proton velocity occurs, the
electromagnetic ion cyclotron (EMIC) instability can drive
preferentially anti-sunward ICWs or the parallel firehose instability
can drive preferentially sunward whistler waves
\citep{Podesta:2011b}.  The growth rate for these instabilities peaks 
at scales $k_\parallel d_i \sim 1$, as shown in
Figure~\ref{fig:unstable}, so these waves are generated at the kinetic
scales associated with the timescale where the parallel signature is
observed and therefore, no parallel cascade is necessary. Our synthetic
spacecraft data models indeed reproduce the features of the observed
magnetic helicity plots when the normalized parallel Poynting flux is
unidirectional, $|S_Z^\parallel|\simeq 1$, and the waves are localized
at parallel wavenumbers in the range $0.3 \lesssim k_\parallel \rho_i$.
We show from a variation of our model parameters that a unidirectional, or nearly
unidirectional, parallel population of waves produces a magnetic helicity
plot which most closely fits with observational results; we do not assume unidirectionallity
based upon the theories presented in \cite{Podesta:2011b}.

Although \citet{Podesta:2011b} demonstrate that anisotropy
instabilities, in the presence of differential alpha-proton drift, can
drive either anti-sunward ICWs or sunward whistlers, they do not
attempt to distinguish between these two alternatives based on 
solar wind data. Here we propose
several observational means to address this issue. 
First, the parallel
firehose instability occurs only for $\beta_{p \parallel}>1$, whereas
the EMIC instability has no such restriction.  Note that, due to the
super-\Alfvenic velocities of the solar wind, the waves measured by a
spacecraft, whether they are anti-sunward or sunward propagating, will
necessarily have been swept past the spacecraft in the anti-sunward
direction.  Therefore, the waves measured by a spacecraft must have
been generated in the solar wind plasma at smaller heliocentric
radius.  Combining this property with the characteristic that the
parallel proton plasma beta, $\beta_{p \parallel}$, in the solar wind
generally increases with heliocentric radius, any waves measured in a
plasma with $\beta_{p \parallel}<1$ will almost certainly have been
generated in a plasma with $\beta_{p\parallel}<1$.  Therefore, if an
interval of solar wind plasma with $\beta_{p\parallel}<1$ contains a
parallel magnetic helicity signature, the parallel firehose
instability cannot have been responsible for the parallel waves.  In
this case, one would have strong evidence that the EMIC instability is
generating anti-sunward ICWs in the solar wind. One cannot, on this
basis, rule out a contribution from the parallel firehose instability
when $\beta_{p \parallel}>1$.

A second way to identify which wave mode is playing a role in the
parallel signature, if the waves are generated locally (in other
words, the waves were generated in the same plasma volume in which
they are measured), is to measure the proton temperature anisotropy of
the plasma. If the temperature anisotropy is of the sense $T_{p
\perp}/T_{p \parallel}>1$, then we expect to find anti-sunward
propagating ICWs caused by the EMIC instability. On the other hand, if
the temperature anisotropy is of the sense $T_{p \perp}/T_{p
\parallel}<1$, then we expect to find sunward propagating whistler
waves caused by the parallel firehose instability. Studies based on
measured distribution functions have shown that the EMIC
instability driven by the proton temperature anisotropy is likely
to operate in high speed streams \citep{Leubner:1986,Marsch:1991,
Marsch:2006,Marsch:2012}. 
Models of the fast solar wind relevant to the \emph{Ulysses} measurements
presented here \citep{Hu:1997,Li:1999} would favor $T_{p \perp}/T_{p \parallel}<1$,
resulting from perpendicular adiabatic cooling of the protons, but direct \emph{in situ}
spacecraft measurements demonstrate that the solar wind plasma has 
intervals with both $T_{p \perp}/T_{p \parallel}<1$ and $T_{p \perp}/T_{p \parallel}>1$ 
\citep{Bale:2009}, although in the ecliptic plane at 1 AU the condition
$T_{p \perp}/T_{p \parallel}<1$ is observed more often than $T_{p \perp}/T_{p \parallel}>1$.

A third way to distinguish anti-sunward ICWs from sunward whistlers,
if electric field measurements are available, is to use electric
field measurements to compute the
field-aligned Poynting flux associated with the parallel signature,
$S_Z^\parallel$.  The convection electric field caused by the flow of
the magnetized plasma past the electric field probes \citep{Chen:2011b}
complicates any effort to compute the Poynting flux, but, in
principle, this approach could conclusively determine which of these
two wave modes is responsible. 
It should be mentioned that
the two instabilities are not necessarily exclusive and that the
solar wind magnetic helicity data is consistent with
the simultaneous existence of both outward propagating ICWs and
inward propagating whistlers. The three observational
tests described above may prove useful in determining if either
one of these instabilities is solely responsible for the quasi-parallel
signature or if both instabilities play a role in its creation.

To conclude the discussion of the waves responsible for the parallel
signature, we consider whether these parallel fluctuations are indeed
turbulent themselves (do they transfer energy nonlinearly to small
scales?), or do they merely persist in the turbulent environment
caused by the anisotropic \Alfvenic fluctuations responsible for the
perpendicular signature. Two properties of the parallel waves argue
against them driving an active turbulent cascade. First, the
amplitudes of the parallel fluctuations are small (since
$P_\perp/P_{tot} \gtrsim 0.9$), so they may not have sufficient
amplitudes for strong nonlinear interactions to occur. If the parallel waves
have larger amplitude but are very intermittent, they still may not
interact frequently enough for nonlinear interactions to play an
important role.  Second, the waves are propagating unidirectionally,
and in the case of incompressible MHD, \Alfven waves traveling in one
direction only do not interact nonlinearly
\citep{Howes:2013a,Nielson:2013}.  Since the linear kinetic waves
are dispersive at $k \rho_i \gtrsim 1$, however, this may enable even
unidirectionally propagating waves to interact nonlinearly.  The
position that the parallel waves do not cascade (at least to smaller
parallel scales) is consistent with the observed short period cutoff of the
parallel signature, especially for the higher frequency \emph{STEREO}
measurements \citep{He:2011a,Podesta:2011a}. 
If the parallel waves are ICWs, this cutoff may 
be due to the efficient damping of the ion cyclotron resonance. However,
if the parallel population is comprised of whistler waves, the cutoff would likely 
not be the result of some linear damping mechanism.
This cutoff appears not
to be due to aliasing, but actually represents the small parallel
scale limit of the parallel waves. The hypothesis that we propose here
is that the parallel waves are indeed not turbulent, but instead
persist alongside the anisotropic \Alfvenic turbulence.  Therefore,
the parallel magnetic helicity signature is a probe of a kinetic
plasma physics mechanisms that operate separately from the turbulent
cascade of energy from large to small scales.


\section{Conclusion}
\label{sec:summary}

Motivated by recent novel measurements of the normalized reduced
fluctuating magnetic helicity in the solar wind, we undertook in this paper to 
illuminate the nature of the underlying turbulence through careful
comparison of \emph{in situ} and synthetic time series.
The parallel and perpendicular signatures which are seen when solar wind
observations of magnetic helicity are segregated by period 
and angle between the local mean magnetic field and solar wind velocity are 
replicated
by synthetic time series derived from physically motivated models
of turbulence. The hypothesis upon which the model is based is 
that the two signatures result from two distinct wavemode populations
in wavevector space, namely a quasi-perpendicular collection of turbulent 
Alfv\'en/kinetic \Alfven
fluctuations and a quasi-parallel collection of either whistler or 
ion cyclotron waves.

In comparing the \emph{in situ} and synthetic magnetic helicity plots,
we have been able to constrain the model's five free parameters,
including the ratio of power contained in quasi-parallel versus quasi-perpendicular 
modes as well as the power ratio of sunward to anti-sunward modes
for each of these two populations. 
From these comparisons, we 
conjecture that the parallel signature is not due to a cascade 
of energy to smaller parallel scales but is more likely due to a 
local injection of parallel energy from temperature anisotropy instabilities.
These parallel modes may not be turbulent, existing along side and having 
little or no interaction with the cascade of energy to smaller 
perpendicular scales.
Magnetic helicity alone is not sufficient to determine the nature
of the parallel fluctuations, but we make several suggestions for 
a future determination of their nature. 
We discuss the effects of 
aliasing and show that the reduction of magnetic helicity amplitude to zero
at small periods,
which has been interpreted as evidence for an increasingly balanced 
turbulent cascade at smaller scales, is likely caused by aliasing and
is not a physical effect. In conclusion, careful comparisons between 
magnetic helicity measurements and synthetic spacecraft data 
is capable of 
providing useful insight into the underlying nature of solar wind turbulence. 

\acknowledgments
This work was supported by NSF CAREER AGS-1054061 and NASA NNX10AC91G.
John Podesta's contribution to this work was supported by the NSF Shine program.
\appendix

\section{Symmetry of the Magnetic Helicity Map about $\theta =90^\circ$}
In this appendix we show that, for angles in the range $0\le \theta \le \pi$, 
the magnetic helicity map derived from 
synthetic spacecraft data is symmetric about the angle $\theta =\pi/2$.  
Consider a spacecraft (or observer) moving with uniform velocity 
$-\V{V}$ along a line oriented at an angle $\theta_0=(\pi/2)-\alpha$ 
with respect to the direction of the constant background field $\V{B}_0$, where 
$|\alpha|\le \pi/2$.  As described in Section 4.1, the magnetic helicity 
spectrum
$\sigma_m^{'}(f_p,\theta)$ is computed using Equations 4 and 5 
from the magnetic field data 
$\V{B}(t_0), \V{B}(t_1), \V{B}(t_2), \ldots, \V{B}(t_{N-1})$  taken as the 
spacecraft moves along that line from point $\V{r}_A$ to point $\V{r}_B$.

Suppose, instead, that the spacecraft moves along the same line but
in the opposite direction 
from point $\V{r}_B$ to point $\V{r}_A$ with velocity $\V{V}$.  
In that case, the observations yield the 
magnetic helicity spectrum at an angle $\theta_1=(\pi/2)+\alpha$.
Thus, the magnetic helicity observed at the angle $\theta_1$ 
is based on those same observations taken at angle $\theta_0$ but in the reverse order: 
$\V{B}(t_{N-1}), \V{B}(t_{N-2}), \V{B}(t_{N-3}), \ldots, \V{B}(t_0)$.  
The only other difference is that the system of basis vectors will be different.  
If $\V{e}_R$, $\V{e}_T$, $\V{e}_N$ are the right-handed 
system of unit 
vectors employed along the trajectory from $\V{r}_A$ to point $\V{r}_B$, 
then along the opposite trajectory from $\V{r}_B$ to $\V{r}_A$ we may use 
the right-handed system of unit vectors $-\V{e}_R$, $\V{e}_T$, 
$-\V{e}_N$.  Therefore, the calculation of the discrete Fourier
transform $\hat B_j(f)$ in Equation 5 for $\theta_0$,
\begin{equation}
\hat B_j(f_p,\theta_0) =\sum_{n=0}^{N-1} B_j(t_n) 
\exp(2\pi i p  n /N),
\end{equation} 
becomes, along the opposite trajectory for $\theta_1$, 
\begin{equation}
\hat B_j(f_p,\theta_1) =\pm \sum_{n=0}^{N-1} B_j(t_{N-1-n}) 
\exp(2\pi i  p n /N),
\label{eqn:theta1}
\end{equation} 
where the minus sign on the right-hand side is used for the components 
$j=R$ and $N$, and the plus sign is used  for the component $j=T$.  
We can rewrite Equation~\ref{eqn:theta1} as 
\begin{equation}
\hat B_j(f_p,\theta_1) =\pm \exp\{2\pi i p (N-1)/N \}
\sum_{n=0}^{N-1} B_j(t_{N-1-n}) 
\exp\{-2\pi i p (N-1-n)/N \}
\end{equation} 
or, equivalently, 
\begin{equation}
\hat B_j(f_p,\theta_1) =\pm \exp\{2\pi i p(N-1)/N \} 
\times [\hat B_j(f_p,\theta_0)]^*,
\end{equation} 
where the asterisk denotes the complex conjugate.  The substitution of this
expression into Equation 4 shows that 
\begin{equation}
\sigma_m^{'}(f_p,\theta_0) =\sigma_m^{'}(f_p,\theta_1).
\end{equation} 
That is, the normalized magnetic helicity spectrum measured along the two 
directions $\theta_0 =(\pi/2)-\alpha$ and $\theta_1 =(\pi/2)+\alpha$ are equal.



\begin{thebibliography}{74}
\expandafter\ifx\csname natexlab\endcsname\relax\def\natexlab#1{#1}\fi

\bibitem[{{Alexandrova} {et~al.}(2009){Alexandrova}, {Saur}, {Lacombe},
  {Mangeney}, {Mitchell}, {Schwartz}, \& {Robert}}]{Alexandrova:2009}
{Alexandrova}, O., {Saur}, J., {Lacombe}, C., {et~al.} 2009, Phys.~Rev.~Lett.,
  103, 165003

\bibitem[{{Bale} {et~al.}(2009){Bale}, {Kasper}, {Howes}, {Quataert}, {Salem},
  \& {Sundkvist}}]{Bale:2009}
{Bale}, S.~D., {Kasper}, J.~C., {Howes}, G.~G., {et~al.} 2009,
  Phys.~Rev.~Lett., 103, 211101

\bibitem[{{Bale} {et~al.}(2005){Bale}, {Kellogg}, {Mozer}, {Horbury}, \&
  {Reme}}]{Bale:2005}
{Bale}, S.~D., {Kellogg}, P.~J., {Mozer}, F.~S., {Horbury}, T.~S., \& {Reme},
  H. 2005, Phys.~Rev.~Lett., 94, 215002

\bibitem[{{Beresnyak} \& {Lazarian}(2008)}]{Beresnyak:2008}
{Beresnyak}, A., \& {Lazarian}, A. 2008, Astrophys.~J., 682, 1070

\bibitem[{{Boldyrev}(2006)}]{Boldyrev:2006}
{Boldyrev}, S. 2006, Phys.~Rev.~Lett., 96, 115002

\bibitem[{{Chandran}(2008)}]{Chandran:2008b}
{Chandran}, B.~D.~G. 2008, Astrophys.~J., 685, 646

\bibitem[{{Chen} {et~al.}(2011){Chen}, {Bale}, {Salem}, \&
  {Mozer}}]{Chen:2011b}
{Chen}, C.~H.~K., {Bale}, S.~D., {Salem}, C., \& {Mozer}, F.~S. 2011,
  Astrophys.~J.~Lett., 737, L41

\bibitem[{{Chen} {et~al.}(2013){Chen}, {Boldyrev}, {Xia}, \&
  {Perez}}]{Chen:2013a}
{Chen}, C.~H.~K., {Boldyrev}, S., {Xia}, Q., \& {Perez}, J.~C. 2013,
  Phys.~Rev.~Lett., 110, 225002

\bibitem[{{Chen} {et~al.}(2010{\natexlab{a}}){Chen}, {Horbury}, {Schekochihin},
  {Wicks}, {Alexandrova}, \& {Mitchell}}]{Chen:2010b}
{Chen}, C.~H.~K., {Horbury}, T.~S., {Schekochihin}, A.~A., {et~al.}
  2010{\natexlab{a}}, Phys.~Rev.~Lett., 104, 255002

\bibitem[{{Chen} {et~al.}(2012){Chen}, {Mallet}, {Schekochihin}, {Horbury},
  {Wicks}, \& {Bale}}]{Chen:2012b}
{Chen}, C.~H.~K., {Mallet}, A., {Schekochihin}, A.~A., {et~al.} 2012,
  Astrophys.~J., 758, 120

\bibitem[{{Chen} {et~al.}(2010{\natexlab{b}}){Chen}, {Wicks}, {Horbury}, \&
  {Schekochihin}}]{Chen:2010a}
{Chen}, C.~H.~K., {Wicks}, R.~T., {Horbury}, T.~S., \& {Schekochihin}, A.~A.
  2010{\natexlab{b}}, Astrophys.~J.~Lett., 711, L79

\bibitem[{{Cho} \& {Lazarian}(2003)}]{Cho:2003a}
{Cho}, J., \& {Lazarian}, A. 2003, Mon.~Not.~Roy.~Astron.~Soc., 345, 325

\bibitem[{{Cho} \& {Vishniac}(2000)}]{Cho:2000b}
{Cho}, J., \& {Vishniac}, E.~T. 2000, Astrophys.~J., 539, 273

\bibitem[{Gary(1986)}]{Gary:1986}
Gary, S.~P. 1986, J.~Plasma Phys., 35, 431

\bibitem[{{Gary} \& {Winske}(1992)}]{Gary:1992}
{Gary}, S.~P., \& {Winske}, D. 1992, J.~Geophys.~Res., 97, 3103

\bibitem[{{Goldreich} \& {Sridhar}(1995)}]{Goldreich:1995}
{Goldreich}, P., \& {Sridhar}, S. 1995, Astrophys.~J., 438, 763

\bibitem[{{Goldstein} {et~al.}(1994){Goldstein}, {Roberts}, \&
  {Fitch}}]{Goldstein:1994}
{Goldstein}, M.~L., {Roberts}, D.~A., \& {Fitch}, C.~A. 1994, J.~Geophys.~Res.,
  99, 11519

\bibitem[{{Hamilton} {et~al.}(2008){Hamilton}, {Smith}, {Vasquez}, \&
  {Leamon}}]{Hamilton:2008}
{Hamilton}, K., {Smith}, C.~W., {Vasquez}, B.~J., \& {Leamon}, R.~J. 2008,
  Journal of Geophysical Research (Space Physics), 113, 1106

\bibitem[{{He} {et~al.}(2011){He}, {Marsch}, {Tu}, {Yao}, \& {Tian}}]{He:2011a}
{He}, J., {Marsch}, E., {Tu}, C., {Yao}, S., \& {Tian}, H. 2011, Astrophys.~J.,
  731, 85

\bibitem[{{He} {et~al.}(2012){He}, {Tu}, {Marsch}, \& {Yao}}]{He:2012}
{He}, J., {Tu}, C., {Marsch}, E., \& {Yao}, S. 2012, Astrophys.~J., 749, 86

\bibitem[{{Hellinger} {et~al.}(2006){Hellinger}, {Tr{\'a}vn{\'{\i}}{\v c}ek},
  {Kasper}, \& {Lazarus}}]{Hellinger:2006}
{Hellinger}, P., {Tr{\'a}vn{\'{\i}}{\v c}ek}, P., {Kasper}, J.~C., \&
  {Lazarus}, A.~J. 2006, Geophys.~Res.~Lett., 33, 9101

\bibitem[{{Higdon}(1984)}]{Higdon:1984}
{Higdon}, J.~C. 1984, Astrophys.~J., 285, 109

\bibitem[{{Horbury} {et~al.}(2008){Horbury}, {Forman}, \&
  {Oughton}}]{Horbury:2008}
{Horbury}, T.~S., {Forman}, M., \& {Oughton}, S. 2008, Phys.~Rev.~Lett., 101,
  175005

\bibitem[{{Horbury} {et~al.}(2012){Horbury}, {Wicks}, \& {Chen}}]{Horbury:2012}
{Horbury}, T.~S., {Wicks}, R.~T., \& {Chen}, C.~H.~K. 2012, ssr, 172, 325

\bibitem[{{Howes}(2008)}]{Howes:2008c}
{Howes}, G.~G. 2008, Phys.~Plasmas, 15, 055904

\bibitem[{{Howes}(2009)}]{Howes:2009}
---. 2009, Nonlin.~Proc.~Geophys., 16, 219

\bibitem[{{Howes}(2011)}]{Howes:2011b}
---. 2011, Astrophys.~J., 738, 40

\bibitem[{{Howes} {et~al.}(2012){Howes}, {Bale}, {Klein}, {Chen}, {Salem}, \&
  {TenBarge}}]{Howes:2012a}
{Howes}, G.~G., {Bale}, S.~D., {Klein}, K.~G., {et~al.} 2012,
  Astrophys.~J.~Lett., 753, L19

\bibitem[{{Howes} {et~al.}(2008{\natexlab{a}}){Howes}, {Cowley}, {Dorland},
  {Hammett}, {Quataert}, \& {Schekochihin}}]{Howes:2008b}
{Howes}, G.~G., {Cowley}, S.~C., {Dorland}, W., {et~al.} 2008{\natexlab{a}},
  J.~Geophys.~Res., 113, 5103

\bibitem[{{Howes} {et~al.}(2008{\natexlab{b}}){Howes}, {Dorland}, {Cowley},
  {Hammett}, {Quataert}, {Schekochihin}, \& {Tatsuno}}]{Howes:2008a}
{Howes}, G.~G., {Dorland}, W., {Cowley}, S.~C., {et~al.} 2008{\natexlab{b}},
  Phys.~Rev.~Lett., 100, 065004

\bibitem[{{Howes} \& {Nielson}(2013)}]{Howes:2013a}
{Howes}, G.~G., \& {Nielson}, K.~D. 2013, Phys.~Plasmas, 20, 072302

\bibitem[{{Howes} \& {Quataert}(2010)}]{Howes:2010a}
{Howes}, G.~G., \& {Quataert}, E. 2010, Astrophys.~J.~Lett., 709, L49

\bibitem[{{Howes} {et~al.}(2011){Howes}, {Tenbarge}, {Dorland}, {Quataert},
  {Schekochihin}, {Numata}, \& {Tatsuno}}]{Howes:2011a}
{Howes}, G.~G., {Tenbarge}, J.~M., {Dorland}, W., {et~al.} 2011,
  Phys.~Rev.~Lett., 107, 035004

\bibitem[{{Hu} {et~al.}(1997){Hu}, {Esser}, \& {Habbal}}]{Hu:1997}
{Hu}, Y.~Q., {Esser}, R., \& {Habbal}, S.~R. 1997, J.~Geophys.~Res., 102, 14661

\bibitem[{{Kasper} {et~al.}(2002){Kasper}, {Lazarus}, \& {Gary}}]{Kasper:2002}
{Kasper}, J.~C., {Lazarus}, A.~J., \& {Gary}, S.~P. 2002, Geophys.~Res.~Lett.,
  29, 170000

\bibitem[{{Kiyani} {et~al.}(2009){Kiyani}, {Chapman}, {Khotyaintsev}, {Dunlop},
  \& {Sahraoui}}]{Kiyani:2009}
{Kiyani}, K.~H., {Chapman}, S.~C., {Khotyaintsev}, Y.~V., {Dunlop}, M.~W., \&
  {Sahraoui}, F. 2009, Phys.~Rev.~Lett., 103, 075006

\bibitem[{{Klein} {et~al.}(2012){Klein}, {Howes}, {TenBarge}, {Bale}, {Chen},
  \& {Salem}}]{Klein:2012}
{Klein}, K.~G., {Howes}, G.~G., {TenBarge}, J.~M., {et~al.} 2012,
  Astrophys.~J., 755, 159

\bibitem[{{Lacombe} \& {Belmont}(1995)}]{Lacombe:1995}
{Lacombe}, C., \& {Belmont}, G. 1995, Advances in Space Research, 15, 329

\bibitem[{{Leamon} {et~al.}(1998){Leamon}, {Matthaeus}, {Smith}, \&
  {Wong}}]{Leamon:1998a}
{Leamon}, R.~J., {Matthaeus}, W.~H., {Smith}, C.~W., \& {Wong}, H.~K. 1998,
  Astrophys.~J.~Lett., 507, L181

\bibitem[{{Leamon} {et~al.}(1999){Leamon}, {Smith}, {Ness}, \&
  {Wong}}]{Leamon:1999}
{Leamon}, R.~J., {Smith}, C.~W., {Ness}, N.~F., \& {Wong}, H.~K. 1999,
  J.~Geophys.~Res., 104, 22331

\bibitem[{{Leubner} \& {Vinas}(1986)}]{Leubner:1986}
{Leubner}, M.~P., \& {Vinas}, A.~F. 1986, J.~Geophys.~Res., 91, 13366

\bibitem[{{Li} {et~al.}(1999){Li}, {Habbal}, {Hollweg}, \& {Esser}}]{Li:1999}
{Li}, X., {Habbal}, S.~R., {Hollweg}, J.~V., \& {Esser}, R. 1999,
  J.~Geophys.~Res., 104, 2521

\bibitem[{{Lithwick} {et~al.}(2007){Lithwick}, {Goldreich}, \&
  {Sridhar}}]{Lithwick:2007}
{Lithwick}, Y., {Goldreich}, P., \& {Sridhar}, S. 2007, Astrophys.~J., 655, 269

\bibitem[{{Maron} \& {Goldreich}(2001)}]{Maron:2001}
{Maron}, J., \& {Goldreich}, P. 2001, Astrophys.~J., 554, 1175

\bibitem[{{Marsch}(1991)}]{Marsch:1991}
{Marsch}, E. 1991, {Kinetic Physics of the Solar Wind Plasma}, ed. R.~{Schwenn}
  \& E.~{Marsch}, 45--133

\bibitem[{{Marsch}(2006)}]{Marsch:2006}
---. 2006, Living Rev.~Solar Phys., 3, 1

\bibitem[{{Marsch}(2012)}]{Marsch:2012}
---. 2012, Space Sci.~Rev., 172, 23

\bibitem[{{Matteini} {et~al.}(2012){Matteini}, {Hellinger}, {Landi},
  {Tr{\'a}vn{\'{\i}}{\v c}ek}, \& {Velli}}]{Matteini:2012}
{Matteini}, L., {Hellinger}, P., {Landi}, S., {Tr{\'a}vn{\'{\i}}{\v c}ek},
  P.~M., \& {Velli}, M. 2012, Space Sci.~Rev., 172, 373

\bibitem[{{Matthaeus} \& {Goldstein}(1982)}]{Matthaeus:1982b}
{Matthaeus}, W.~H., \& {Goldstein}, M.~L. 1982, J.~Geophys.~Res., 87, 6011

\bibitem[{{Matthaeus} {et~al.}(1982){Matthaeus}, {Goldstein}, \&
  {Smith}}]{Matthaeus:1982a}
{Matthaeus}, W.~H., {Goldstein}, M.~L., \& {Smith}, C. 1982, Phys.~Rev.~Lett.,
  48, 1256

\bibitem[{{Narita} {et~al.}(2011){Narita}, {Gary}, {Saito}, {Glassmeier}, \&
  {Motschmann}}]{Narita:2011}
{Narita}, Y., {Gary}, S.~P., {Saito}, S., {Glassmeier}, K.-H., \& {Motschmann},
  U. 2011, Geophys.~Res.~Lett., 38, 5101

\bibitem[{{Nielson} {et~al.}(2013){Nielson}, {Howes}, \&
  {Dorland}}]{Nielson:2013}
{Nielson}, K.~D., {Howes}, G.~G., \& {Dorland}, W. 2013, Phys.~Plasmas, 20,
  072303

\bibitem[{{Osman} {et~al.}(2012){Osman}, {Matthaeus}, {Hnat}, \&
  {Chapman}}]{Osman:2012}
{Osman}, K.~T., {Matthaeus}, W.~H., {Hnat}, B., \& {Chapman}, S.~C. 2012,
  Phys.~Rev.~Lett., 108, 261103

\bibitem[{{Perez} \& {Boldyrev}(2009)}]{Perez:2009}
{Perez}, J.~C., \& {Boldyrev}, S. 2009, Phys.~Rev.~Lett., 102, 025003

\bibitem[{{Perri} {et~al.}(2012){Perri}, {Goldstein}, {Dorelli}, \&
  {Sahraoui}}]{Perri:2012}
{Perri}, S., {Goldstein}, M.~L., {Dorelli}, J.~C., \& {Sahraoui}, F. 2012,
  Phys.~Rev.~Lett., 109, 191101

\bibitem[{{Podesta}(2009)}]{Podesta:2009}
{Podesta}, J.~J. 2009, Astrophys.~J., 698, 986

\bibitem[{{Podesta}(2013)}]{Podesta:2013}
---. 2013, Sol.~Phys., 286, 529

\bibitem[{{Podesta} \& {Gary}(2011{\natexlab{a}})}]{Podesta:2011b}
{Podesta}, J.~J., \& {Gary}, S.~P. 2011{\natexlab{a}}, Astrophys.~J., 742, 41

\bibitem[{{Podesta} \& {Gary}(2011{\natexlab{b}})}]{Podesta:2011a}
---. 2011{\natexlab{b}}, Astrophys.~J., 734, 15

\bibitem[{{Quataert}(1998)}]{Quataert:1998}
{Quataert}, E. 1998, Astrophys.~J., 500, 978

\bibitem[{{Roberts} {et~al.}(2013){Roberts}, {Li}, \& {Li}}]{Roberts:2013}
{Roberts}, O.~W., {Li}, X., \& {Li}, B. 2013, Astrophys.~J., 769, 58

\bibitem[{{Sahraoui} {et~al.}(2010){Sahraoui}, {Goldstein}, {Belmont}, {Canu},
  \& {Rezeau}}]{Sahraoui:2010}
{Sahraoui}, F., {Goldstein}, M.~L., {Belmont}, G., {Canu}, P., \& {Rezeau}, L.
  2010, Phys.~Rev.~Lett., 105, 131101

\bibitem[{{Sahraoui} {et~al.}(2009){Sahraoui}, {Goldstein}, {Robert}, \&
  {Khotyaintsev}}]{Sahraoui:2009}
{Sahraoui}, F., {Goldstein}, M.~L., {Robert}, P., \& {Khotyaintsev}, Y.~V.
  2009, Phys.~Rev.~Lett., 102, 231102

\bibitem[{{Salem} {et~al.}(2012){Salem}, {Howes}, {Sundkvist}, {Bale},
  {Chaston}, {Chen}, \& {Mozer}}]{Salem:2012}
{Salem}, C.~S., {Howes}, G.~G., {Sundkvist}, D., {et~al.} 2012,
  Astrophys.~J.~Lett., 745, L9

\bibitem[{{Schekochihin} {et~al.}(2009){Schekochihin}, {Cowley}, {Dorland},
  {Hammett}, {Howes}, {Quataert}, \& {Tatsuno}}]{Schekochihin:2009}
{Schekochihin}, A.~A., {Cowley}, S.~C., {Dorland}, W., {et~al.} 2009,
  Astrophys.~J.~Supp., 182, 310

\bibitem[{{Song} {et~al.}(1994){Song}, {Russell}, \& {Gary}}]{Song:1994}
{Song}, P., {Russell}, C.~T., \& {Gary}, S.~P. 1994, J.~Geophys.~Res., 99, 6011

\bibitem[{{Stix}(1992)}]{Stix:1992}
{Stix}, T.~H. 1992, {Waves in plasmas}

\bibitem[{{Taylor}(1938)}]{Taylor:1938}
{Taylor}, G.~I. 1938, Royal Society of London Proceedings Series A, 164, 476

\bibitem[{{TenBarge} \& {Howes}(2012)}]{TenBarge:2012a}
{TenBarge}, J.~M., \& {Howes}, G.~G. 2012, Phys.~Plasmas, 19, 055901

\bibitem[{{TenBarge} \& {Howes}(2013)}]{TenBarge:2013a}
---. 2013, Astrophys.~J.~Lett., 771, L27

\bibitem[{{TenBarge} {et~al.}(2013){TenBarge}, {Howes}, \&
  {Dorland}}]{TenBarge:2013c}
{TenBarge}, J.~M., {Howes}, G.~G., \& {Dorland}, W. 2013, Astrophys.~J., in
  press

\bibitem[{{Verdon} {et~al.}(2009){Verdon}, {Cairns}, {Melrose}, \&
  {Robinson}}]{Verdon:2009}
{Verdon}, A.~L., {Cairns}, I.~H., {Melrose}, D.~B., \& {Robinson}, P.~A. 2009,
  Phys.~Plasmas, 16, 052105

\bibitem[{{Verscharen} {et~al.}(2013){Verscharen}, {Bourouaine}, \&
  {Chandran}}]{Verscharen:2013b}
{Verscharen}, D., {Bourouaine}, S., \& {Chandran}, B.~D.~G. 2013, ArXiv
  e-prints

\bibitem[{{Wicks} {et~al.}(2010){Wicks}, {Horbury}, {Chen}, \&
  {Schekochihin}}]{Wicks:2010}
{Wicks}, R.~T., {Horbury}, T.~S., {Chen}, C.~H.~K., \& {Schekochihin}, A.~A.
  2010, Mon.~Not.~Roy.~Astron.~Soc., 407, L31

\end{thebibliography}

\end{document}